%% file: main.tex
\pdfoutput=1
\documentclass[12pt,a4paper]{article}

\usepackage{adjustbox}
\usepackage{bm}
\usepackage{dcolumn}

\usepackage{ifthen} 
\newboolean{pdflatex}
\setboolean{pdflatex}{true} 

\newboolean{articletitles}
\setboolean{articletitles}{true} 

\newboolean{uprightparticles}
\setboolean{uprightparticles}{false} 

\def\paperauthors{LHCb collaboration} 
\def\paperasciititle{Lc polarimetry using the dominant hadronic mode} 
\def\papertitle{\Lc polarimetry using the dominant hadronic mode} 
\def\paperkeywords{{High Energy Physics}, {LHCb}} 
\def\papercopyright{\the\year\ CERN for the benefit of the LHCb collaboration} 
\def\paperlicence{CC BY 4.0 licence}
\def\paperlicenceurl{https://creativecommons.org/licenses/by/4.0/}

\input{preamble}
\usepackage{longtable} 

\begin{document}

\newcommand{\modified}[1]{{\color{black}#1}}
\newcommand{\unpolrate}{\ensuremath{I_0}\xspace} 
\newcommand{\diff}{\mathrm{d}}

\renewcommand{\thefootnote}{\fnsymbol{footnote}}
\setcounter{footnote}{1}

\input{title-LHCb-PAPER}

\renewcommand{\thefootnote}{\arabic{footnote}}
\setcounter{footnote}{0}

\cleardoublepage

\pagestyle{plain} 
\setcounter{page}{1}
\pagenumbering{arabic}


\input{body}

\subsection*{Code Availability Statement} 
A software used for this paper is available at {\sc{Zenodo}} repository~\cite{polarimetry.COMPWA:2022xyz}.
The analysis documentation is hosted at the webpage, \href{https://lc2pkpi-polarimetry.docs.cern.ch/}{lc2pkpi-polarimetry.docs.cern.ch}.

\input{acknowledgements}

\input{supplementary}

\input{supplementary-app}

\addcontentsline{toc}{section}{References}
\input{references.bbl}

\newpage
\newcommand{\lhcborcid}[1]{\href{https://orcid.org/#1}{\hspace*{0.1em}\raisebox{-0.45ex}{\includegraphics[width=1em]{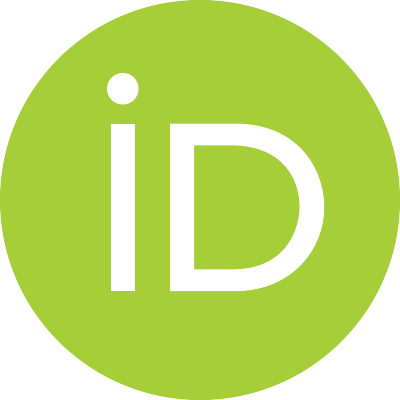}}}}
\input{Authorship_LHCb-PAPER-2022-044}

\end{document}

%% file: preamble.tex
\usepackage[top=1in, bottom=1.25in, left=1in, right=1in]{geometry}

\usepackage{microtype}
\usepackage{lineno}  
\usepackage{xspace} 
\usepackage{caption} 

\usepackage{graphicx}  
\usepackage{color}
\usepackage{colortbl}

\usepackage{amsmath} 
\usepackage{amssymb}
\usepackage{amsfonts}
\usepackage{upgreek} 

\newcommand*\patchAmsMathEnvironmentForLineno[1]{%
\expandafter\let\csname old#1\expandafter\endcsname\csname #1\endcsname
\expandafter\let\csname oldend#1\expandafter\endcsname\csname
end#1\endcsname
 \renewenvironment{#1}%
   {\linenomath\csname old#1\endcsname}%
   {\csname oldend#1\endcsname\endlinenomath}%
}
\newcommand*\patchBothAmsMathEnvironmentsForLineno[1]{%
  \patchAmsMathEnvironmentForLineno{#1}%
  \patchAmsMathEnvironmentForLineno{#1*}%
}
\AtBeginDocument{%
\patchBothAmsMathEnvironmentsForLineno{equation}%
\patchBothAmsMathEnvironmentsForLineno{align}%
\patchBothAmsMathEnvironmentsForLineno{flalign}%
\patchBothAmsMathEnvironmentsForLineno{alignat}%
\patchBothAmsMathEnvironmentsForLineno{gather}%
\patchBothAmsMathEnvironmentsForLineno{multline}%
\patchBothAmsMathEnvironmentsForLineno{eqnarray}%
}

\usepackage{hyperxmp}

\usepackage[pdftex,
            pdfauthor={\paperauthors},
            pdftitle={\paperasciititle},
            pdfkeywords={\paperkeywords},
            pdfcopyright={Copyright (C) \papercopyright},
            pdflicenseurl={\paperlicenceurl}]{hyperref}

\usepackage[bottom,flushmargin,hang,multiple]{footmisc}

\usepackage[all]{hypcap} 

\input{lhcb-symbols-def} 

\usepackage{cite} 
\usepackage{mciteplus}

%% file: lhcb-symbols-def.tex
\usepackage{xspace} 
\usepackage{upgreek}

\def\lhcb   {\mbox{LHCb}\xspace}

\def\MagUp {\mbox{\em Mag\kern -0.05em Up}\xspace}

\ifthenelse{\boolean{uprightparticles}}%
{

 \def\Pnu         {\ensuremath{\upnu}\xspace}                 
                  
 \def\Ppi         {\ensuremath{\uppi}\xspace}

 \def\Ptau        {\ensuremath{\uptau}\xspace}

 \def\Ppsi        {\ensuremath{\uppsi}\xspace}

 \def\PDelta      {\ensuremath{\Delta}\xspace}                 
 \def\PXi         {\ensuremath{\Xi}\xspace}                 
 \def\PLambda     {\ensuremath{\Lambda}\xspace}                 
 \def\PSigma      {\ensuremath{\Sigma}\xspace}                 
 \def\POmega      {\ensuremath{\Omega}\xspace}                 
 \def\PUpsilon    {\ensuremath{\Upsilon}\xspace}
 \let\oldPi\Pi
 \def\PPi         {\ensuremath{\oldPi}\xspace}

 \def\PB      {\ensuremath{\mathrm{B}}\xspace}                 
                  
 \def\PD      {\ensuremath{\mathrm{D}}\xspace}

 \def\PJ      {\ensuremath{\mathrm{J}}\xspace}                 
 \def\PK      {\ensuremath{\mathrm{K}}\xspace}

 \def\Pb      {\ensuremath{\mathrm{b}}\xspace}                 
 \def\Pc      {\ensuremath{\mathrm{c}}\xspace}                 
                  
 \def\Pe      {\ensuremath{\mathrm{e}}\xspace}

 \def\Pi      {\ensuremath{\mathrm{i}}\xspace}

 \def\Pp      {\ensuremath{\mathrm{p}}\xspace}

 \def\Ps      {\ensuremath{\mathrm{s}}\xspace}

 \def\thebaroffset{0.0em}
}
{

 \def\Pnu         {\ensuremath{\nu}\xspace}                 
                  
 \def\Ppi         {\ensuremath{\pi}\xspace}

 \def\Ptau        {\ensuremath{\tau}\xspace}

 \def\Ppsi        {\ensuremath{\psi}\xspace}                 
                  
 \mathchardef\PDelta="7101
 \mathchardef\PXi="7104
 \mathchardef\PLambda="7103
 \mathchardef\PSigma="7106
 \mathchardef\POmega="710A
 \mathchardef\PUpsilon="7107
 \mathchardef\PPi="7105
                  
 \def\PB      {\ensuremath{B}\xspace}                 
                  
 \def\PD      {\ensuremath{D}\xspace}

 \def\PJ      {\ensuremath{J}\xspace}                 
 \def\PK      {\ensuremath{K}\xspace}

 \def\Pb      {\ensuremath{b}\xspace}                 
 \def\Pc      {\ensuremath{c}\xspace}                 
                  
 \def\Pe      {\ensuremath{e}\xspace}

 \def\Pi      {\ensuremath{i}\xspace}

 \def\Pp      {\ensuremath{p}\xspace}

 \def\Ps      {\ensuremath{s}\xspace}

 \def\thebaroffset{0.18em}
}
\newcommand{\offsetoverline}[2][\thebaroffset]{\kern #1\overline{\kern -#1 #2}}%

\makeatletter
\ifcase \@ptsize \relax
  \newcommand{\miniscule}{\@setfontsize\miniscule{4}{5}}
\or
  \newcommand{\miniscule}{\@setfontsize\miniscule{5}{6}}
\or
  \newcommand{\miniscule}{\@setfontsize\miniscule{5}{6}}
\fi
\makeatother

\DeclareRobustCommand{\optbar}[1]{\shortstack{{\miniscule (\rule[.5ex]{1.25em}{.18mm})}
  \\ [-.7ex] $#1$}}

\def\en         {{\ensuremath{\Pe^-}}\xspace}   
\def\ep         {{\ensuremath{\Pe^+}}\xspace}

\def\taum       {{\ensuremath{\Ptau^-}}\xspace}

\def\ellm       {{\ensuremath{\ell^-}}\xspace}
\def\ellp       {{\ensuremath{\ell^+}}\xspace}

\def\neub       {{\ensuremath{\overline{\Pnu}}}\xspace}

\def\neulb      {{\ensuremath{\neub_\ell}}\xspace}

\def\squark    {{\ensuremath{\Ps}}\xspace}

\def\cquark    {{\ensuremath{\Pc}}\xspace}

\def\bquark    {{\ensuremath{\Pb}}\xspace}

\def\pion   {{\ensuremath{\Ppi}}\xspace}

\def\pip    {{\ensuremath{\pion^+}}\xspace}
\def\pim    {{\ensuremath{\pion^-}}\xspace}

\def\kaon    {{\ensuremath{\PK}}\xspace}

\def\KorKbar {\kern \thebaroffset\optbar{\kern -\thebaroffset \PK}{}\xspace}

\def\Kp      {{\ensuremath{\kaon^+}}\xspace}
\def\Km      {{\ensuremath{\kaon^-}}\xspace}

\def\D       {{\ensuremath{\PD}}\xspace}

\def\DorDbar {\kern \thebaroffset\optbar{\kern -\thebaroffset \PD}\xspace}

\def\Dp      {{\ensuremath{\D^+}}\xspace}
\def\Dm      {{\ensuremath{\D^-}}\xspace}

\def\DpDm    {\ensuremath{\Dp {\kern -0.16em \Dm}}\xspace}

\def\B       {{\ensuremath{\PB}}\xspace}

\def\BorBbar {\kern \thebaroffset\optbar{\kern -\thebaroffset \PB}\xspace}

\def\Bd      {{\ensuremath{\B^0}}\xspace}

\def\BdorBdbar {\kern \thebaroffset\optbar{\kern -\thebaroffset \Bd}\xspace}
\def\Bu      {{\ensuremath{\B^+}}\xspace}

\def\Bp      {{\ensuremath{\Bu}}\xspace}

\def\Bs      {{\ensuremath{\B^0_\squark}}\xspace}

\def\BsorBsbar {\kern \thebaroffset\optbar{\kern -\thebaroffset \Bs}\xspace}

\def\jpsi     {{\ensuremath{{\PJ\mskip -3mu/\mskip -2mu\Ppsi}}}\xspace}

\def\Y#1S{\ensuremath{\PUpsilon{(#1S)}}\xspace}

\def\proton      {{\ensuremath{\Pp}}\xspace}

\def\Deltares    {{\ensuremath{\PDelta}}\xspace}

\def\Lz          {{\ensuremath{\PLambda}}\xspace}
\def\Lbar        {{\ensuremath{\offsetoverline{\PLambda}}}\xspace}
\def\LorLbar     {\kern \thebaroffset\optbar{\kern -\thebaroffset \PLambda}\xspace}
\def\Lambdares   {{\ensuremath{\PLambda}}\xspace}

\def\Xires       {{\ensuremath{\PXi}}\xspace}

\def\Omegares    {{\ensuremath{\POmega}}\xspace}

\def\Lc          {{\ensuremath{\Lz^+_\cquark}}\xspace}
\def\Lcbar       {{\ensuremath{\Lbar{}^-_\cquark}}\xspace}
\def\Xic         {{\ensuremath{\Xires_\cquark}}\xspace}

\def\Xicp        {{\ensuremath{\Xires^+_\cquark}}\xspace}

\def\Omegac      {{\ensuremath{\Omegares^0_\cquark}}\xspace}

\def\Lb           {{\ensuremath{\Lz^0_\bquark}}\xspace}

\def\Omegab       {{\ensuremath{\Omegares^-_\bquark}}\xspace}

\newcommand{\decay}[2]{\ensuremath{#1\!\to #2}\xspace} 

\def\to                 {\ensuremath{\rightarrow}\xspace}

\def\CP                {{\ensuremath{C\!P}}\xspace}

\def\AT#1     {\ensuremath{A_{\mathrm{T}}^{#1}}\xspace}           

\def\C#1      {\ensuremath{\mathcal{C}_{#1}}\xspace}                       
\def\Cp#1     {\ensuremath{\mathcal{C}_{#1}^{'}}\xspace}                    
\def\Ceff#1   {\ensuremath{\mathcal{C}_{#1}^{\mathrm{(eff)}}}\xspace}        
\def\Cpeff#1  {\ensuremath{\mathcal{C}_{#1}^{'\mathrm{(eff)}}}\xspace}       
\def\Ope#1    {\ensuremath{\mathcal{O}_{#1}}\xspace}                       
\def\Opep#1   {\ensuremath{\mathcal{O}_{#1}^{'}}\xspace}                    

\newcommand{\aunit}[1]{\ensuremath{\text{\,#1}}}       

\newcommand{\tev}{\aunit{Te\kern -0.1em V}\xspace}
\newcommand{\gev}{\aunit{Ge\kern -0.1em V}\xspace}
\newcommand{\mev}{\aunit{Me\kern -0.1em V}\xspace}
\newcommand{\kev}{\aunit{ke\kern -0.1em V}\xspace}
\newcommand{\ev}{\aunit{e\kern -0.1em V}\xspace}
 
\newcommand{\mevc}{\ensuremath{\aunit{Me\kern -0.1em V\!/}c}\xspace}
\newcommand{\gevc}{\ensuremath{\aunit{Ge\kern -0.1em V\!/}c}\xspace}
\newcommand{\mevcc}{\ensuremath{\aunit{Me\kern -0.1em V\!/}c^2}\xspace}
\newcommand{\gevcc}{\ensuremath{\aunit{Ge\kern -0.1em V\!/}c^2}\xspace}

\def\fb   {\ensuremath{\aunit{fb}}\xspace}
\def\invfb   {\ensuremath{\fb^{-1}}\xspace}

\def\gsim{{~\raise.15em\hbox{$>$}\kern-.85em
          \lower.35em\hbox{$\sim$}~}\xspace}
\def\lsim{{~\raise.15em\hbox{$<$}\kern-.85em
          \lower.35em\hbox{$\sim$}~}\xspace}

\def\tell1  {TELL1\xspace}
\def\ukl1   {UKL1\xspace}

\newcommand{\eg}{\mbox{\itshape e.g.}\xspace}
\newcommand{\ie}{\mbox{\itshape i.e.}\xspace}



%% file: title-LHCb-PAPER.tex

\begin{titlepage}
\pagenumbering{roman}

\vspace*{-1.5cm}
\centerline{\large EUROPEAN ORGANIZATION FOR NUCLEAR RESEARCH (CERN)}
\vspace*{1.5cm}
\noindent
\begin{tabular*}{\linewidth}{lc@{\extracolsep{\fill}}r@{\extracolsep{0pt}}}
\ifthenelse{\boolean{pdflatex}}
{\vspace*{-1.5cm}\mbox{\!\!\!\includegraphics[width=.14\textwidth]{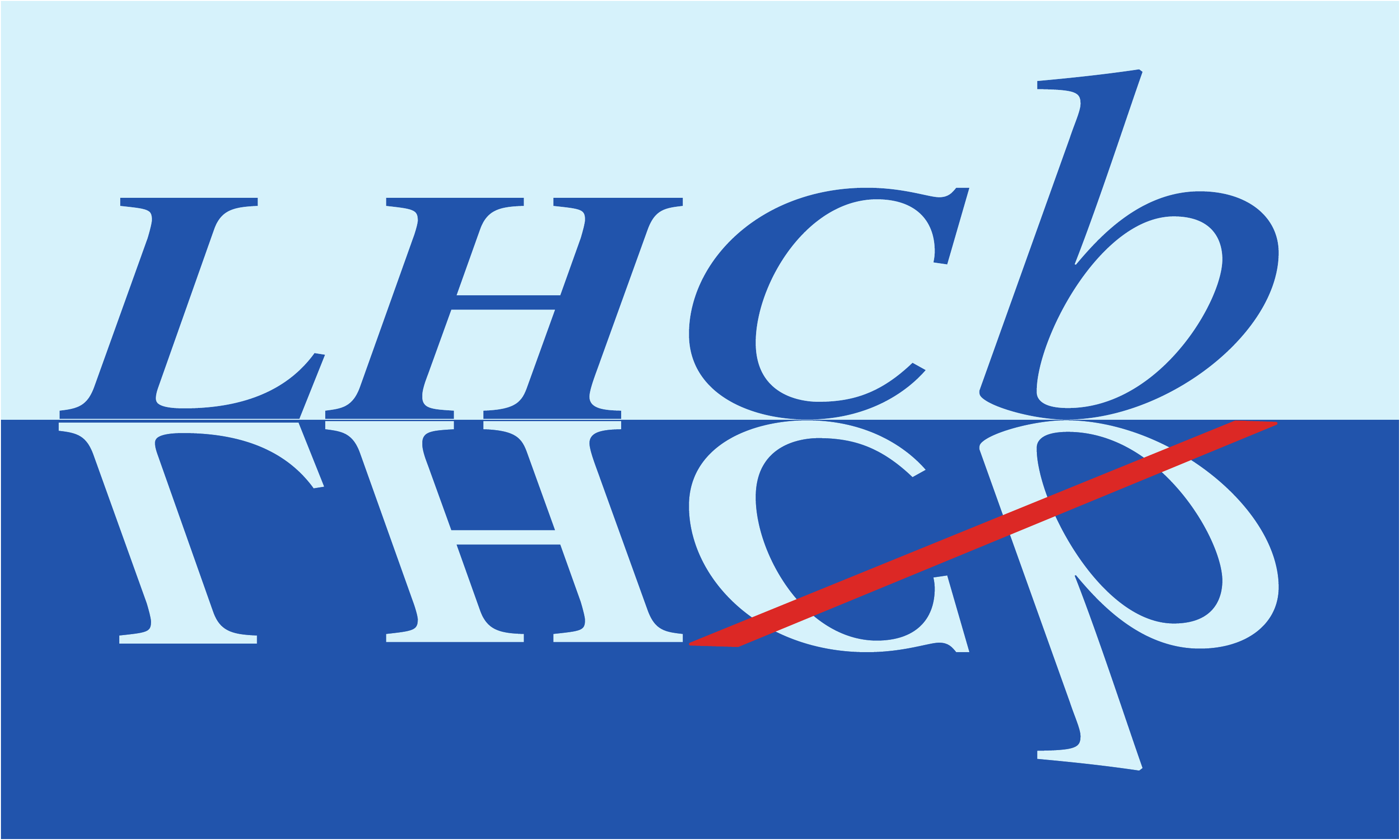}} & &}%
{\vspace*{-1.2cm}\mbox{\!\!\!\includegraphics[width=.12\textwidth]{figs/lhcb-logo.eps}} & &}%
\\
 & & CERN-EP-2022-287 \\  
 & & LHCb-PAPER-2022-044 \\  
 & & \today \\ 
 & & \\
\end{tabular*}

\vspace*{4.0cm}

{\normalfont\bfseries\boldmath\huge
\begin{center}
  \papertitle 
\end{center}
}

\vspace*{2.0cm}

\begin{center}
\paperauthors\footnote{Authors are listed at the end of this paper.}
\end{center}

\vspace{\fill}

\begin{abstract}
  \noindent
%
The polarimeter vector field for multibody decays of a spin-half baryon is introduced as a generalisation of the baryon asymmetry parameters.
Using a recent amplitude analysis of the $\Lc \to p \Km \pip$ decay performed at the LHCb experiment, we compute the distribution of the kinematic-dependent polarimeter vector for this process in the space of Mandelstam variables to express the polarised decay rate in a model-agnostic form.
The obtained representation can facilitate polarisation measurements of the $\Lc$ baryon and eases inclusion of the~$\Lc \to p \Km \pip$ decay mode in hadronic amplitude analyses.
\end{abstract}

\vspace*{1.0cm}

\begin{center}
  Published in JHEP 07 (2023) 228
\end{center}

\vspace{\fill}

{\footnotesize 
\centerline{\copyright~\papercopyright. \href{\paperlicenceurl}{\paperlicence}.}}
\vspace*{2mm}

\end{titlepage}


\newpage
\setcounter{page}{2}
\mbox{~}

%% file: body.tex
\section{Introduction}

Polarisation characterises a preferred orientation of the particle spin in space, and it is conserved for a freely moving particle.
Therefore, the strength and direction of the polarisation of particles produced in collisions or decays gives access to details of the underlying fundamental theory.
Polarisation is widely exploited in spectroscopy studies of conventional and exotic QCD states.
Particularly, studies of pentaquarks in $\Lb\to P_\cquark^+(\to \jpsi p) \Km$ decays in the LHCb experiment
take advantage of angular correlations of the muon pairs appearing in the $\jpsi\to \mu^+\mu^-$ transition,
which improves the sensitivity to the pentaquark spin~\cite{LHCb-PAPER-2015-029}.%
\footnote{The charge-conjugate is implied for reactions and particles throughout the text, unless stated otherwise.}
Angular analysis of the quantum numbers of
$\Omegares_\cquark^{**0}$ baryon states,%
\footnote{The double-star superscript is used throughout the paper to indicate all excited states of the hadron families.}
appearing as $\Xicp \Km$ resonances in $\Omegab \to \Xicp \pi^- \Km$ decays, is found to be sensitive to the $\Omegares_\cquark^{**0}$ baryon parity only when the angular distribution of the $\Xicp \to \proton \Km \pip$ transition is considered~\cite{LHCb-PAPER-2021-012}.
This, in turn, requires detailed knowledge of the $\Xicp \to \proton \Km \pip$ decay matrix elements.
The polarisation of promptly produced mesons and baryons is an essential observable that is sensitive to the mechanisms behind quark hadronisation~\cite{Brambilla:2010cs, Faccioli:2010kd, Butenschoen:2012px}.
Many proposed searches for physics beyond the Standard Model are based on semileptonic decays of the $\Lb$ baryon, such as $\Lb \to \Lc \ellm \neulb$ transitions~\cite{Konig:1993wz,Dutta:2015ueb,Shivashankara:2015cta,Li:2016pdv,Datta:2017aue,Ray:2018hrx,DiSalvo:2018ngq,Penalva:2019rgt,Ferrillo:2019owd}, where
certain properties of the $b\to c$ transition current are reflected in the $\Lc$ polarisation.
For instance, the sign of the longitudinal polarisation of the $\Lc$ baryon in $\Lb$ decays provides a test for the left-handedness of the $b\to c$ current~\cite{Konig:1993wz}. The transition amplitude of the $\Lc \to \proton \Km \pip$ decay recently determined by \lhcb~\cite{LHCb-PAPER-2022-002} enables such measurement of the polarisation.
A new opportunity to test the Standard Model by measuring the electric and magnetic dipole moment of charm baryons using spin rotation in crystals at the LHC experiments is discussed in Refs.~\cite{Baryshevsky:2016cul,Botella:2016ksl,Fomin:2017ltw}.
Recent feasibility studies~\cite{Fomin:2019wuw,Aiola:2020yam} show that the measurement of the electric dipole moment of the $\Lc$ and $\Xicp$ baryons can be performed in the near future.

In general, the measurement of the polarisation in baryon decays deserves special consideration,
since the final state of the decay contains a baryon, \eg a proton, whose spin state is usually not observed.
Due to the averaging over the baryon spin state, the influence of the initial-state polarisation on the measurable observables is diminished.
For example, angular distributions in the decay of a fermion to a fermion and a scalar are insensitive to the initial polarisation of the fermion when the decay conserves parity.
However, the forward-backward asymmetry in the distribution of the helicity angle of the decay products is non-zero in weak decays, where both the parity-conserving and parity-violating currents contribute to the transition amplitude. In that case, the differential decay rate reads~\cite{PDG2022}
\begin{align} \label{eq:P.alpha.cos}
    \frac{2}{\Gamma}\frac{\diff \Gamma}{\diff\cos\theta} = 1 + P \alpha \cos\theta\,,
\end{align}
where $\Gamma$ is the partial width for the considered decay, $\theta$ is the helicity angle of the decay-product fermion, $P$ is the longitudinal polarisation of the initial state, and $\alpha$ is the parity-violating \textit{asymmetry parameter} of the reaction.
This parameter serves two roles.
First, it is a fundamental hadronic observable that characterises parity violation in the weak transition. 
Second, it is a practical quantity, since the value of $\alpha$ does not depend on the production mechanism of the decaying fermion.
The measurement of the angular distributions in the decay is a universal tool to study the physics behind the production polarisation once its asymmetry parameter is known.

The observables for polarisation measurements have been extensively explored for the decays of the $\taum$ lepton~\cite{Tsai:1971vv, Kuhn:1991cc, Davier:1992nw}. For the polarised decay, the differential decay rate reads
\begin{align} \label{eq:diff.rate.tau}
    \frac{\Phi}{\Gamma}\frac{\diff \Gamma}{\mathrm{d}\Phi} \propto 1+ \vec{P}\cdot\vec{h}\,,
\end{align}
where $\Phi$ is the Lorentz-invariant phase space for the decay,
$\vec{P}$ is the polarisation vector of the $\taum$ lepton in its rest frame, and $\vec{h}$ is a \textit{polarimeter vector}.
In semileptonic decays, the polarimeter vector is a unit vector whose direction depends on the final state~\cite{Kuhn:1995nn, Kuhn:1982di, Kuhn:1993ra, Hagiwara:1989fn, Davier:1992nw}.

For the decay of a spin-half hadron, Eq.~\eqref{eq:diff.rate.tau} still holds, but the length of the polarimeter vector can be less than~$1$.
The direction of $\vec h$ is related to the orientation of the final state particles, but does not have to coincide with any of the momentum vectors.
Instead, the length and orientation of the polarimeter vector depends on the kinematic variables of the decay.
As shown in this paper, the dependence on the decay-plane orientation can be factored out as
\begin{align} \label{eq:h.via.alpha}
    \vec{h} = R(\phi,\theta,\chi) \vec{\alpha}\,,
\end{align}
where $R$ is a three-dimensional rotation matrix with
$\phi$, $\theta$, and $\chi$ being the Euler angles that describe the orientation of the decay products in space.
The \textit{aligned polarimeter vector}, denoted by $\vec\alpha$, describes the direction of the polarimeter vector with respect to the decay-product vectors. The dependence of $\vec{\alpha}$ on the kinematic variables is specific to the decay reaction and needs to be determined experimentally.
For two-body decays with a baryon and a pseudoscalar in the final state, the vector $\vec{\alpha}$ only has a longitudinal component, which matches the asymmetry parameter~$\alpha$ in Eq.~\eqref{eq:P.alpha.cos}.
For multibody decays, the $\vec{\alpha}$ vector forms a continuous vector field in the space of the kinematic variables.


The three-body decay of $\Lc \to \proton \Km \pip$ is the golden hadronic mode for reconstructing this charm baryon experimentally. Firstly, this decay mode has one of the largest branching fraction among all hadronic $\Lc$ decays.
Secondly, the final state of this decay only consists of charged particles, which have a high reconstruction efficiency.
The transition amplitude of the $\Lc \to \proton \Km \pip$ decays was studied previously by the ACCMOR collaboration~\cite{Jezabek:1992ke, Jezabek:1992vi} and the E791 collaboration~\cite{E791:1999ajq,Fox:1999ja}. 
These studies indicated a large polarisation in the production of the $\Lc$ baryon, but a detailed amplitude description could not be obtained due to the limited statistics and difficulties in formulating the amplitude model.
A new result on the amplitude analysis has been published by the \lhcb collaboration recently, based on a sample of $400\,000$ decays with an equal amount of $\Lc$ and $\Lcbar$ baryons~\cite{LHCb-PAPER-2022-002}.
The candidates are selected from semileptonic decays of beauty hadrons produced in $pp$ collisions at a centre-of-mass energy of $13\tev$, corresponding to an integrated luminosity of $1.7\invfb$.\footnote{Natural units with $c = 1$ are used throughout this paper.}
Further information about the experimental setup relevant for the present studies can be found in Ref.~\cite{LHCb-PAPER-2022-002}.
The study reveals the presence of twelve different decay chains with resonances in all two-body subsystems:
six $\Lambdares^{**}$ states, three $\Deltares^{**++}$ states, and three $K^{**0}$ states.
The average polarisation of the $\Lc$ baryon in semileptonic decays of beauty hadrons is also measured.
The asymmetry parameter $\alpha$ is reported for quasi-two-body decays of the $\Lc$ state to a baryon and a pseudoscalar.

The utilisation of the $\Lc\to p\Km\pip$ decay parametrisation is highly non-trivial given the complexity and model-dependence of the amplitude analysis.
In this paper, we simplify the usage of these results by converting the transition amplitude for the $\Lc\to \proton \Km \pip$ decay to a model-agnostic representation using the distribution of the aligned polarimeter vector $\vec{\alpha}$.
The components of the vector are computed on a two-dimensional grid of the Dalitz-plot variables.
Special attention is devoted to the propagation of uncertainties.
The theoretical aspects of the present paper are in line with the original consideration of the $\Lc$ polarimetry in three-body decays done by Bjorken~\cite{Bjorken:1988ya}.
The recent work of Wei~\textit{et al.}~\cite{Wei:2022kem} discusses the $\Lc\to p\Km \pip$ decay from a similar angle,
by suggesting measuring the model-independent parameters referred to as the \textit{calibration parameters}.
We find that these parameters are equal to the averaged polarimeter components.
The values are reported in this document for completeness. We find a significant reduction of the uncertainty on the polarisation measurement if the complete polarimetry field is used instead of the averaged values.

The paper is structured as follows. In Section~\ref{sec:alpha.multibody}, we discuss general features of the polarisation measurement in multibody decays and introduce the polarimeter vector in general form by factoring out the overall rotation.
Section~\ref{sec:Lc2pKpi} is dedicated to the computation of the polarimeter distribution for $\Lc \to \proton \Km \pip$ decays, and we conclude in Section~\ref{sec:conclusion}.
Appendix~\ref{sec:DPD} presents the amplitude model.
The relation between the model parameters used in this analysis and in the previous analysis~\cite{LHCb-PAPER-2022-002} are given in Appendix~\ref{sec:LHCb.mapping}.
The \CP violation effects in the $\Lc$ decays are discussed in Appendix~\ref{sec:CP}.
Appendix~\ref{sec:example} illustrates how to use the polarimetry fields in spectroscopy studies by writing the reaction amplitude for $B^+ \to \Lc \Lcbar K^+$ decays with all 12 kinematic variables.


\section{Polarisation sensitivity} \label{sec:alpha.multibody}

We consider a multibody decay of a spin-half fermion decaying to three or more particles.
The total differential decay rate $\diff \Gamma / \diff\Phi$ for the polarised decay of this fermion is computed as $\left|\mathcal{M}\right|^2 / (2m_0)$,
where $m_0$ is the mass of the decaying particle, and $\left|\mathcal{M}\right|^2$ is the averaged squared matrix element.
The latter is obtained by combining the transition amplitudes with their conjugates through the spin-density matrix and summing over all spin indices,
\begin{align} \label{eq:I.tot}
    \left|\mathcal{M}\right|^2 = \sum_{\nu_0,\nu_0',\{\lambda\}}\mathfrak{R}_{\nu_0',\nu_0} T_{\nu_0',\{\lambda\}}^* T_{\nu_0,\{\lambda\}}\,,
\end{align}
where $T_{\nu_0^{(\prime)},\{\lambda\}}$ is the transition amplitude,
$\nu_0^{(\prime)}$ is the spin projection of the decaying fermion to the $z$~axis of the production plane, and $\{\lambda\}  = \{\lambda_1, \lambda_2, \dots, \lambda_n\}$ is the combined index for the helicities of the final state particles numbered by the low index $i$.
The spin-density matrix $\mathfrak{R}$ is a two-dimensional matrix in the spin-projection basis computed in the rest frame of the decaying particle, with three polarisation degrees of freedom, $\mathfrak{R}_{\nu_0',\nu_0} = 1+\vec P \cdot \vec \sigma^P_{\nu_0',\nu_0}$,
where \mbox{$\vec P = (P_x,P_y,P_z)$} is a polarisation vector and
\mbox{$\vec\sigma^P = (\sigma^P_x,\sigma^P_y,\sigma^P_z)$} are the Pauli matrices.
The coordinate system for the polarisation vector and the spin quantisation axis are fixed by the production process. The $z$~axis is given by the momentum of the decay particle, the $y$~axis is normal to the production plane, and the $x$~axis completes the right-handed coordinate system.

\begin{figure}
    \centering
    \includegraphics[width=0.7\textwidth]{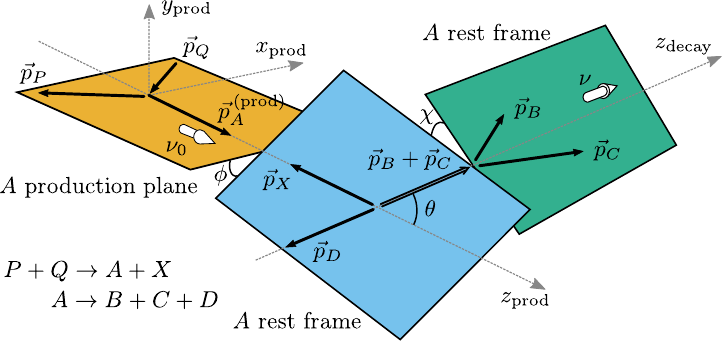}
    \caption{Definition of the decay-plane orientation angles $\phi$, $\theta$, and $\chi$ related to polarisation of the particle $A$, produced in the process $P+Q \to A+X$, and observed in the decay, $A\to B+C+D$. The left yellow plane contains the momenta of the production process,
    while the right green plane contains the momenta of the decay reaction.
    The central blue plane is defined by the momenta $\vec{p}_X$ and $\vec{p}_D$ in the rest frame of the decay particle $A$.
    The white arrows show the orientation of the two different quantisation axes of the particle $A$ spin in its rest frame.}
    \label{fig:angles}
\end{figure}

The rotation properties of the transition amplitude $T_{\nu_0,\{\lambda\}}$ are defined by the total spin of the system, which is $1/2$ for a fermion decay.
For every point in the decay phase space, the configuration of the final-state particle momenta forms a rigid body that can be rotated to the common \textit{aligned configuration}~\cite{JPAC:2019ufm}, where one of the momenta is chosen as reference for the $z$~axis and the second noncollinear vector is used to establish the direction perpendicular to the decay plane, the \mbox{$y$~axis}.
The entire angular dependence of the decay amplitude is factored out explicitly,
\begin{align} \label{eq:T.gen.DPD}
    T_{\nu_0,\{\lambda\}} = \sum_{\nu} D_{\nu_0,\nu}^{1/2*}(\phi,\theta,\chi) A_{\nu,\{\lambda\}}(\kappa)\,,
\end{align}
where 
the index~$\nu$ is the spin projection of the decaying fermion onto the \mbox{$z$~axis} of the aligned configuration,
$D_{\nu_0,\nu}^{1/2}$ is the Wigner~$D$-function, $A_{\nu,\{\lambda\}}$ is the transition amplitude in the aligned configuration,
and $\kappa$ denotes the kinematic variables which the aligned reaction amplitude depends on.
The number of variables is counted as $3n-7$, with $n$ being the number of the decay products, and the seven constraints originate from the four conservation laws and the three overall rotations of the system. Notably, the Euler angles in the \mbox{$Z$-$Y$-$Z$} convention shown in Fig.~\ref{fig:angles} coincide with the regular helicity angles used in the construction of the decay amplitude as a sequence of two-body transitions.
Namely, for a process $P+Q \to A(\to B+C+D)\,+X$, the angles $(\theta, \phi)$, and $\chi$ are the spherical angles of the momentum sum $\vec{p}_B+\vec{p}_C$ in the helicity frame of $A$ obtained from the centre-of-momentum frame of the reaction, and the polar angle of particle $B$ in the helicity frame of $BC$ obtained from the $A$ rest frame, respectively.

By substituting Eq.~\eqref{eq:T.gen.DPD} into Eq.~\eqref{eq:I.tot},
we get an expression for the polarised decay rate in terms of the kinematic variables and the rotational degrees of freedom,
\begin{align} \label{eq:intensity.full}
    \left|\mathcal{M}\right|^2 = \sum_{\nu_0,\nu_0',\nu,\nu'} \mathfrak{R}_{\nu_0',\nu_0}\,\,
     D_{\nu_0',\nu'}^{1/2*}(\phi,\theta,\chi)\, D_{\nu_0,\nu}^{1/2}(\phi,\theta,\chi)\,\,
     \mathfrak{X}_{\nu',\nu}(\kappa)\,,
\end{align}
where 
$\mathfrak{X}$ is a hermitian $2\times 2$ matrix that encapsulates the entire dependence on the decay dynamics.
It can be expanded in the basis of the Pauli matrices and the identity matrix,
\begin{align} \label{eq:polarimetrty.matrix}
    \mathfrak{X}_{\nu',\nu}(\kappa) &= \sum_{\{\lambda\}} A_{\nu',\{\lambda\}}^*(\kappa) A_{\nu,\{\lambda\}}(\kappa)\,,\\ \nonumber
    &=\frac{I_0(\kappa)}{2} \left(1+\vec\alpha(\kappa) \cdot \vec \sigma^{P}\right)_{\nu',\nu}\,.
\end{align}
Here, $I_0(\kappa)$ is the total differential decay rate, \mbox{$I_0 = \sum_{\nu,\{\lambda\}} \left|A_{\nu,\{\lambda\}} \right|^2$}.
The aligned polarimeter vector $\vec\alpha$ is computed by expanding the squared decay amplitude in the basis of the Pauli matrices,
\begin{align} \label{eq:alpha.def}
    \vec\alpha(\kappa) =  \sum_{\nu',\nu,\{\lambda\}} A^{*}_{\nu',\{\lambda\}}\vec\sigma_{\nu',\nu}  A_{\nu,\{\lambda\}} \,\big /\, \unpolrate(\kappa)\,.
\end{align}

Equation~\eqref{eq:intensity.full} is simplified using properties of the rotation group and the Pauli matrices.
The Wigner~$D$-matrix in Eq.~\eqref{eq:T.gen.DPD} has $4\pi$ invariance, since it belongs to the \mbox{spin-$1/2$} representation of the rotation group.
However, dependence of the physical observables on the rotation must correspond to the physical, \mbox{$2\pi$-folded} representations.
Particularly, the Cornwell theorem from group theory (see for example Section 3, Chapter 5 of Ref.~\cite{Cornwell:1997ke}) gives the relation between the rotation of the transition amplitude (\mbox{spin-$1/2$} representation of the SU(2) group) and the rotation of the three-dimensional vector (spin-1 representation of the SU(2) group). %
With this, the expression for the differential decay rate gets a simple form,
\begin{align} \label{eq:master.intensity}
    \left|\mathcal{M}(\phi, \theta, \chi, \kappa)\right|^2 =
        \unpolrate(\kappa)\, \bigg(1 + \sum_{i,j} P_i R_{ij}(\phi, \theta, \chi) \alpha_j(\kappa) \bigg)\,,
\end{align}
where $R_{ij}(\phi,\theta,\chi)$ is a three-dimensional rotation matrix implementing the Euler transformation to a physical vector.
It matches Eq.~\eqref{eq:diff.rate.tau} with the matrix product, $R\,\vec{\alpha}$, being the polarimeter vector~$\vec{h}$.
The quantities $\vec\alpha(\kappa)$ give a model-agnostic representation for polarisation dependence of the decay rate.
To incorporate the decay variables to a more complex reaction, Eq.~\eqref{eq:polarimetrty.matrix} should be used.
For example, an amplitude analysis of $\Bp\to \Lc\Lcbar \Kp$, where \Lc and \Lcbar decays to $p \Km\pip$ and $\bar{p} \Kp \pim$ final states, respectively,
would greatly benefit from the polarimeter field~$\vec\alpha$. The decay rate is expressed using the $\mathfrak{X}^\Lc$ and $\mathfrak{X}^\Lcbar$ matrices in Appendix~\ref{sec:example}.

The differential decay rate remains sensitive to the initial polarisation even after
integration over the kinematic variables $\kappa$.
In that case, the rate is only a function of the three production angles,
\begin{align} \label{eq:I.alpha.averaged} 
    \frac{8\pi^2}{\Gamma}\frac{\diff^3 \Gamma}{\diff\phi\,\diff\cos\theta\,\diff\chi} = 1+\sum_{i,j}P_i R_{ij}(\phi, \theta, \chi) \overline{\alpha}_j\,,
\end{align}
where the components of the \textit{averaged aligned polarimeter vector} $\vec{\overline{\alpha}}$ are defined as,
\begin{align} \label{eq:alpha.averaged}
    \overline{\alpha}_j = \int  \unpolrate\alpha_j \diff^n \kappa \, \big/\, \int  \unpolrate\,\diff^n \kappa\,.
\end{align}
The integration over the kinematic variables simplifies the analysis,
but leads to an increased uncertainty on the results.
As discussed below, the method proposed in Ref.~\cite{Davier:1992nw} can be used to quantify this effect.


The polarisation measurement for $\taum$ lepton decays is formulated in general terms in Refs.~\cite{Rouge:1991pm, Davier:1992nw}.
The authors introduce a \textit{scalar polarisation sensitivity} $S = 1/(\sigma\sqrt{N})$, where $N$ is the number of events in the sample analysed using the unbinned likelihood method and $\sigma$ is the expected uncertainty for the value of the polarisation.
The method has been applied to evaluate the scalar polarisation sensitivity of the $\Lc\to p\Km\pip$ decays in Refs.~\cite{Botella:2016ksl, LHCb-PAPER-2022-002}.
If the probability density function for the decay rate reads as $W = f + p g$,
with $f$ and $g$ being the normalised polarisation-independent and polarisation-dependent terms of the decay rate, respectively,
and $p$ being a small component of the polarisation vector,
then
the squared polarisation sensitivity~$S^2$, computed as the second derivative of the negative log-likelihood, is equal to the variance of the ratio, $g / f$~\cite{Davier:1992nw}.
Using Eq.~\eqref{eq:master.intensity}, it is straightforward to identify the $f$ and $g$ functions:
$f = \unpolrate(\kappa) / f_0$, and $g = \sum_j \unpolrate(\kappa) R_{3j}(\phi, \theta, \chi) \alpha_j(\kappa) / f_0$, where $f_0$ is the normalisation constant.
For small polarisation values, the variance can be evaluated for $p=0$, so that the three integrals over the Euler angles are analytic.
\begin{align}  \nonumber
    S_0^2 &= \int \left.\frac{g^2}{f+ p\,g} \right|_{p=0}  \diff^{3} (\phi, \theta, \chi)\,\diff^n\kappa\\ 
    &= 
    \frac{1}{3} \int \unpolrate \left|\vec{\alpha}\right|^2 \diff^n \kappa \,\big /\, \int \unpolrate\,\diff^n \kappa\,,
    \label{eq:s0.integrals}
\end{align}
where the factor~$3$ reflects the vector properties of~$\vec{\alpha}$.
Equation~\eqref{eq:s0.integrals} has a simple interpretation: the polarisation sensitivity is proportional to the averaged length of the polarimeter vector over the phase space domain, weighted by the probability distribution for the unpolarised decay.
When an analysis only uses angular variables, as in Eq.~\eqref{eq:I.alpha.averaged},
the polarisation sensitivity can be computed from the length of the averaged polarimeter vector with
\mbox{$\overline{S}_0^2 = (\overline{\alpha}_x^2+\overline{\alpha}_y^2+\overline{\alpha}_z^2)/3$}.
The ratio $S_0 / \overline{S}_0$ gives the expected increase of the statistical uncertainty on the polarisation value if the kinematic dependence is ignored.

\section{Polarisation sensitivity for \texorpdfstring{$\boldsymbol{\Lc \to p \Km \pip }$}{Λc⁺ → pK⁻π⁺} decays}\label{sec:Lc2pKpi}

A three-body decay is the smallest system where the polarimetry field develops a non-trivial behaviour.
In this case, the aligned amplitude depends on two kinematic variables.
The natural choice for these degrees of freedom are the Mandelstam variables, often presented in a Dalitz plot~\cite{Dalitz:1953cp},
in which the aligned polarimeter vector forms a continuous vector field.

To compute the polarisation sensitivity for the \mbox{$\Lc\to p \Km \pip $} decays, the results of the recent amplitude analysis of this decay in Ref.~\cite{LHCb-PAPER-2022-002} are used.
The model of Ref.~\cite{LHCb-PAPER-2022-002} is mapped to the conventions used in the present analysis in Appendix~\ref{sec:LHCb.mapping}.
There are two important differences in our setup.
First, the decay plane in the $\Lc$ rest frame is oriented using the $K^{**0}$ decay chains as a reference: 
the sum of the pion and kaon momenta defines the positive $z$~direction and the frame is oriented such that the $x$~component of the pion momentum is positive.
For the previous analysis, the proton is aligned with the $z$~axis and the $x$~component of the pion momentum is negative, \ie all momenta vectors are rotated by $\pi$ about the $y$~axis.
Second, the alignment of the decay chains is performed for the helicity state of the proton by matching the rotations as worked out in Ref.~\cite{JPAC:2019ufm} and the alignment rotations are expressed via the Thomas angle (see derivation on p.~215 of Ref.~\cite{Gourgoulhon:2013gua}) for the canonical state of the proton in the model of Ref.~\cite{LHCb-PAPER-2022-002}.
The couplings are defined to match the two ways of formulating the amplitude.
The obtained model is validated against the implementation of Ref.~\cite{LHCb-PAPER-2022-002} up to differences in floating-point precision.

\begin{figure}
    \centering
    \includegraphics[width=0.75\linewidth]{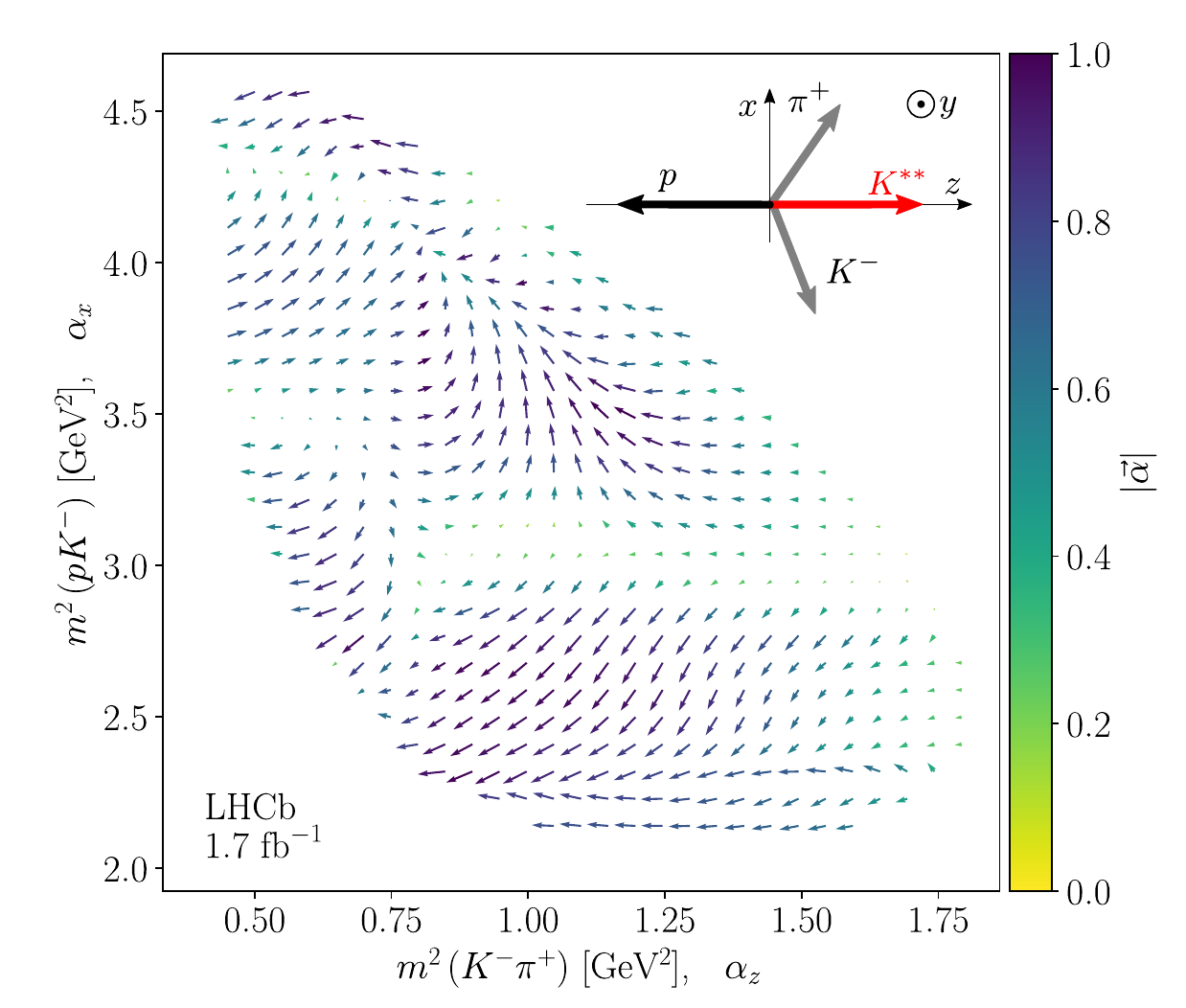}%
    \caption{Aligned polarimeter vector field in Dalitz-plot coordinates.
    The $z$ and $x$ components of the $\alpha$~vector are shown by the horizontal and vertical projections of the arrow, respectively. The colour indicates the length of the polarimeter vector.
    The sketch in the top right corner shows the decay-plane orientation.
    The momentum arrows for the pion and the kaon are shown in gray, since their orientation depends on the kinematic variables, $m^2(\Km\pip)$ and $m^2(p \Km)$.
    }
    \label{fig:alpha}
\end{figure}

\begin{figure}
    \centering
    \includegraphics[width=0.49\linewidth]{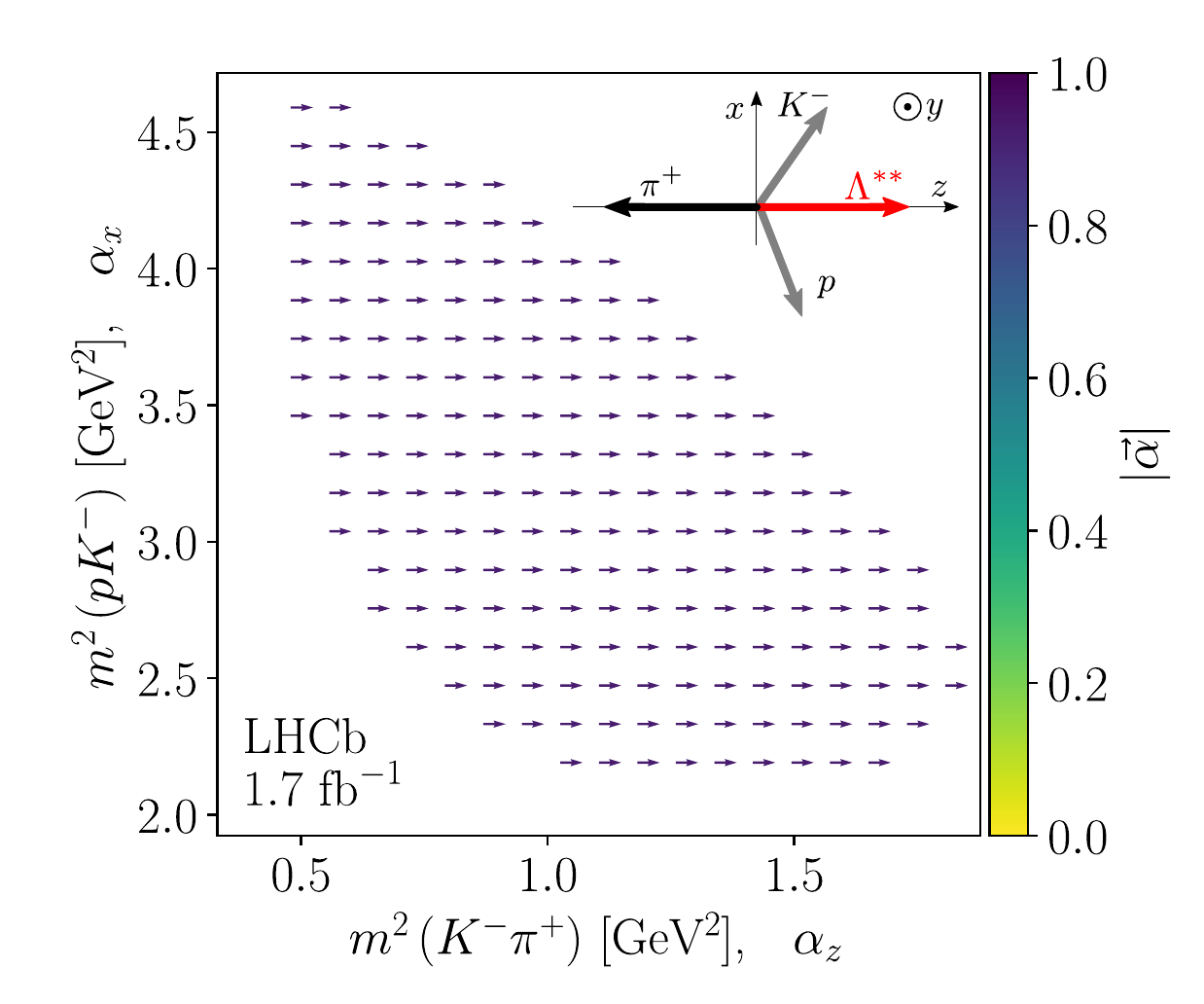}
    \includegraphics[width=0.49\linewidth]{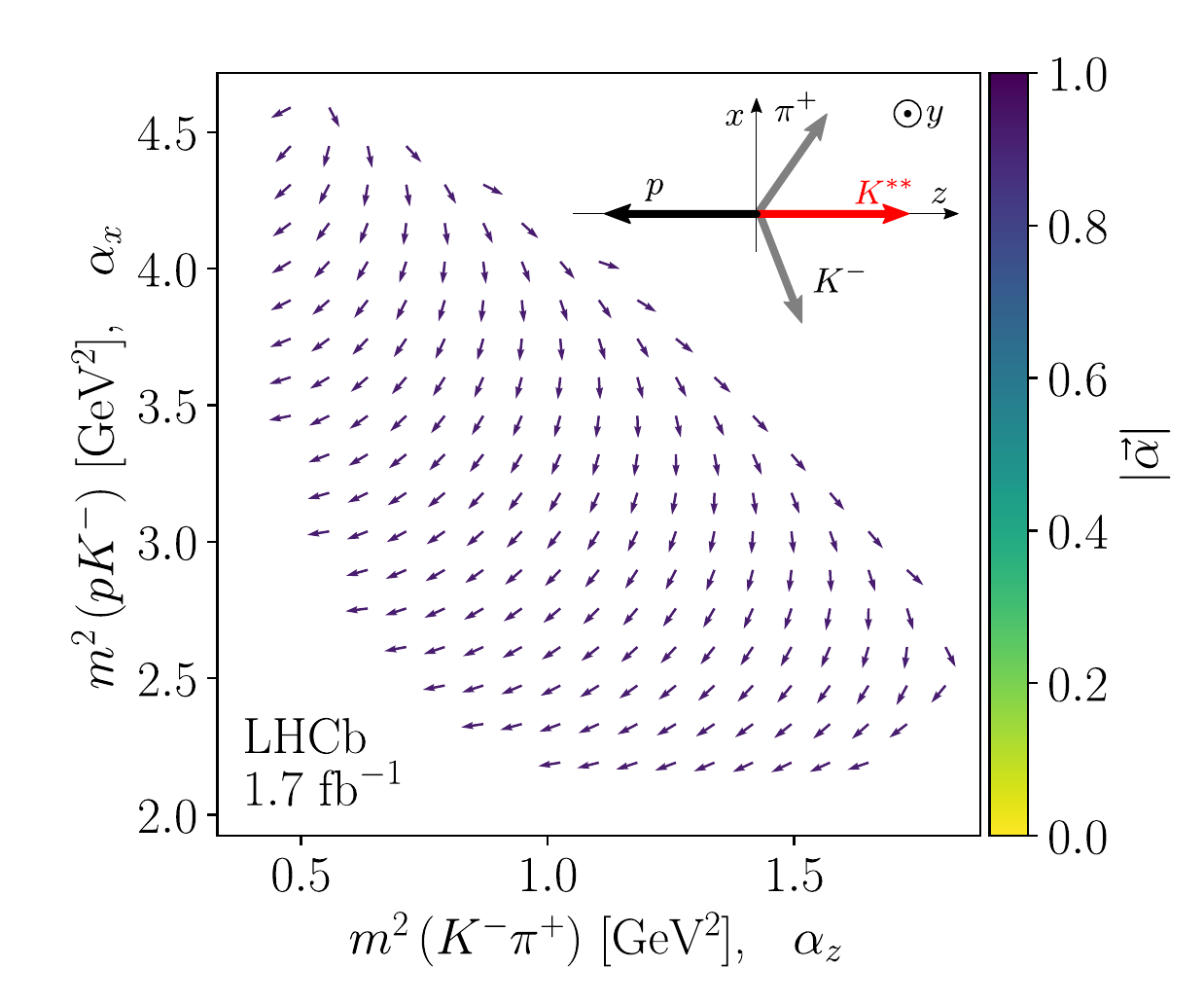}
    \caption{The aligned polarimeter vector field in Dalitz-plot coordinates for the \mbox{$\Lc \to \Lambdares(1520)(\to p K^-) \pi^+$} decay chain only.
    The decay plane in the rest frame of the $\Lc$ baryon is aligned to the $K^{**0}$ decay chains and $\Lambdares^{**}$ decay chain for the left and right panel, respectively. See Fig.~\ref{fig:alpha} for notations.} 
    \label{fig.alignment}
\end{figure}

The distribution of the $\vec\alpha$~components computed using Eq.~\eqref{eq:alpha.def} is shown in Fig.~\ref{fig:alpha}.
Together with the intensity distribution, the $\vec\alpha$~field gives all information needed to determine the \Lc polarisation in further analyses by means of Eq.~\eqref{eq:master.intensity}.
The polarimeter vector at the phase-space point given in the Dalitz-plot coordinates,  $m^2(\Km\pip)$ and $m^2(\Km p)$, is indicated with an arrow projected onto the \mbox{$xz$~plane}.
The length of the polarimeter vector, shown by the colour, changes from point to point. However, it is greater than $0.5$ for most of the kinematic domain,
indicating significant contributions of both parity-conserving and parity-violating currents~\cite{Dedu:2742640}.
The structures in Fig.~\ref{fig:alpha} are driven by resonances in different subsystems and their interference.
For the $\Lc$ baryon decaying to a baryon and a (pseudo)scalar, the aligned polarimeter vector points in the same direction as the momentum of one of the two particles.
The resonance then decays,
but this does not influence the direction of the polarimeter vector.
If the $z$~axis is chosen parallel to the momentum of the resonance of the decay chain,
the vector map is homogeneous, \ie the polarimeter vector is the same at every point in the phase space, as shown on the left panel of Fig.~\ref{fig.alignment}.
For a combination of many decay chains, $\vec\alpha$ is not aligned with any momenta.
In fact, interference between the decay chains might cause $\alpha_y \neq 0$, meaning that the \mbox{$\vec\alpha$~vector} points out of the momentum plane. 

If an alternative aligned configuration is obtained from the default one by the rotation $R$, the $\alpha$~vector is transformed under the inverse rotation,
\begin{align} \label{eq:alpha.prime}
    \alpha_i \xrightarrow[R]{} \alpha_i' = \sum_{j} R^{-1}_{ij} \alpha_j\,,
\end{align}
where $\vec{\alpha}^{\,\prime}$ is the aligned polarimeter vector for the transformed configuration.
The default alignment choice in this paper is $\vec{z} \uparrow\uparrow -\vec{p}_p$, \ie the proton momentum is opposite to the $z$ axis.
To update the convention for either $\vec{z} \uparrow \uparrow -\vec p_{\pip}$ or $\vec{z} \uparrow \uparrow -\vec p_{\Km}$, one needs to apply $R_y^{-1}(\zeta^0_{i(1)})$ for every point of the $\vec\alpha$~field, where $i=2$ or $i=3$, respectively.
The angle $\zeta^0_{i(1)}$ is the Wigner angle for spin alignment of the $\Lc$ baryon, as given in~\cite{JPAC:2019ufm}.
Figure~\ref{fig.alignment} shows the $\vec\alpha$~field for the $\Lc \to \Lambdares(1520) \pi^+$ decay chain with two different decay-plane alignments.
For this decay chain, the aligned polarimeter vector always points along the momentum of the $\Lambdares(1520)$ resonance. When the proton momentum is used as a reference for the $z$ axis, the polarimeter vector is tilted for every point on the map.
The rotation $R$ in Eq.~\eqref{eq:alpha.prime} that relates the alternative conventions for the Euler angles is global, \ie it does not depend on the kinematic variables.
For example, in Ref.~\cite{LHCb-PAPER-2022-002} the $z$~axis in the aligned configuration is determined by the momentum of the proton, rather than the opposite direction, as done here. With this convention, we get $R = R_y(\pi)$, which flips the sign of $\alpha_x$ and $\alpha_z$ for every point of the Dalitz plot. 
By using the alternative particle ordering, $p(1) K^-(2) \pi^+(3)$, it can be seen that the flip of the decay plane corresponds to the global $R_z(\pi)$ rotation.
In the conventions of Ref.~\cite{Wei:2022kem}, the proton momentum determines the $x$ axis
with $xy$ being the decay plane, and the kaon momentum having the positive $y$ projection.
This aligned configuration is obtained with the rotation $R_{W} = R_z(\pi/2)R_y(\pi/2)$.

The uncertainty on the results of the amplitude analysis consists of three classes:
the statistical uncertainty, related to the size of the data sample, the model uncertainty, inferred from a number of alternative setups, and the systematic uncertainty, which originates from experimental unknowns.
These uncertainties are propagated to the computed polarimeter vector fields and other inferred quantities.

The statistical uncertainties are computed using the corresponding uncertainty on the parameters for the default model of Ref.~\cite{LHCb-PAPER-2022-002}. The parameter values are assumed to be Gaussian-distributed and uncorrelated.
A sample of 100 parameter sets is drawn from the model parameters to compute the distribution of the inferred quantities.
To preserve correlations of the polarimeter field in the kinematic domain, the sampled $\vec\alpha$ distributions are provided as supplementary material~\cite{polarimetry.COMPWA:2022xyz}.

The model uncertainties are computed from the alternative models that are described in Ref.~\cite{LHCb-PAPER-2022-002}.
The polarimeter field is computed for every alternative model and presented in supplemental material for this paper.
For scalar inferred quantities, the model uncertainty is computed by the extrema of the sample of alternative models.
These model uncertainties dominate over the statistical uncertainties for all the checked quantities.
Once the $\vec\alpha$ field is used for the polarisation measurements with Eq.~\eqref{eq:master.intensity},
or for a spectroscopy analysis (see Appendix~\ref{sec:example} for an example),
it becomes straightforward to propagate all uncertainties by simply replacing the field of the default model to the fields provided in the supplementary material.

\begin{figure}
    \centering
    \includegraphics[width=\linewidth]{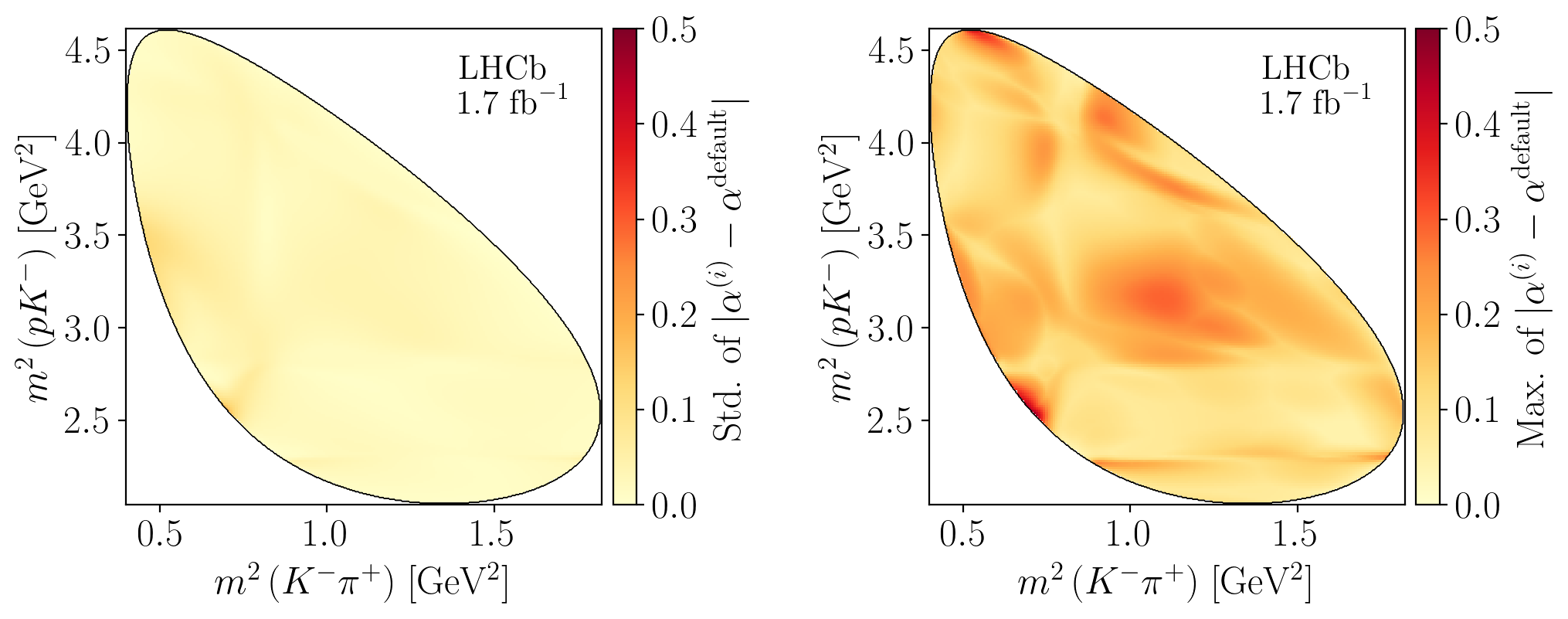}
    \caption{Uncertainties on the length of the aligned polarimeter vector for each phase-space point. The left panel shows combined statistical and systematic uncertainties, and the right panel shows the model uncertainties.}
    \label{fig:alpha-uncertainties}
\end{figure}

The systematic uncertainty is found to be negligible compared to the statistical and model uncertainties.
The effect is accounted for by inflating the statistical uncertainties used in the resampling, namely,
the statistic and systematic uncertainties on the sampled parameters are added in quadrature.
Figure~\ref{fig:alpha-uncertainties} summarises the effect of the uncertainties on the polarimetry field.
The model uncertainty dominates over the other sources of uncertainty. Areas with the largest uncertainties overlap with regions where there is a small number of events and where tails of different resonances interfere.
All computations are performed with libraries from the ComPWA project\cite{Fritsch:2022compwa} and cross-checked with the \texttt{ThreeBodyDecay.jl} package\cite{Mikhasenko:2022xyz}.


A distinct feature of multibody decays with respect to two-body processes is that the averaged value of the polarimeter vector defined by Eq.~\eqref{eq:alpha.averaged} has transverse components.
The averaged asymmetry
$\vec{\overline{\alpha}}$
is computed with Eq.~\eqref{eq:I.alpha.averaged}.
The result for the averaged aligned polarimeter vector is shown in Table~\ref{tab:averaged.alphas}.
\newcolumntype{d}[1]{D{.}{.}{#1}}
\begin{table}[]
    \caption{Averaged values of the polarimetry fields in Cartesian and spherical coordinates. The central value and the model uncertainties are shown in the second and fourth columns, respectively. The second column presents the standard deviation of the distribution in the parameter sampling that includes both statistical and systematic uncertainties.}
    \centering
    \begin{tabular}
    {l r | d{3.2} d{3.2} d{3.2}}
    \multicolumn{2}{c|}{Observable} & \multicolumn{1}{c}{Central} & \multicolumn{1}{c}{Stat. and syst.} & \multicolumn{1}{c}{Model} \\ \hline
    $\overline{\alpha}_x$ & [$\%$] &  -6.26  &  0.45  & 1.48  \\
    $\overline{\alpha}_y$ & [$\%$] &  0.89   &  0.89  & 1.27  \\
    $\overline{\alpha}_z$ & [$\%$] & -27.80  &  2.37  & 4.04  \\
    \hline
    $\overline{\alpha}$ & [$\%$]      & 28.51  &  2.40 & 3.79  \\ 
    $\overline{\theta}$ & [rad]       & 2.92  & <0.01 & 0.05  \\ 
    $\overline{\phi}$   & [rad]       & 3.00  &  0.14 & 0.21  \\
    \end{tabular}
    \label{tab:averaged.alphas}
\end{table}
In addition to the Cartesian components, the table gives the averaged vector in spherical coordinates, with $\overline{\alpha}$, $\overline{\theta}$, and $\overline{\phi}$ being
the length of the averaged polarimeter vector and its polar and azimuthal angles, respectively.
Correlation coefficients in the spherical representation are expected to be lower.
Indeed, we find the correlation over the uncertainties on the $(\overline{\alpha}_z,\overline{\alpha}_x)$ components to be about $100\%$, and the correlation coefficient of $(\overline{\alpha}_y, \overline{\alpha}_x)$, $(\overline{\alpha}_z, \overline{\alpha}_y)$ to be both about $10\%$.
The correlation of the spherical components is found to be $70\%$, $50\%$, and $10\%$ for
$(\overline{\alpha},\overline{\theta})$,
$(\overline{\theta},\overline{\phi})$, and
$(\overline{\alpha},\overline{\phi})$, respectively.
The non-zero value of $\overline{\alpha}_y$ is caused by the interference of various decay chains contributing to the $\Lc$ decays, since it vanishes for all chains considered individually.
The polarisation sensitivity $\sqrt{3}\;\overline{S}_0$ with the averaged polarimeter vector is found to be $0.29\pm 0.02\pm0.04$.
It can be compared to $\sqrt{3}\,S_0 = 0.67\pm0.01\pm0.02$ when the polarimetry field is used.
Hence, the polarisation can also be measured using the integrated values from Eq.~\eqref{eq:I.alpha.averaged}
at the cost of an increased statistical uncertainty on the polarisation by, roughly, a factor~$2.35$.
The polarisation uncertainty due to the uncertainty on $\vec \alpha$ shown in Fig.~\ref{fig:alpha-uncertainties} can be estimated from the uncertainties on $S_0$.
As such, one expects that the model uncertainty on the polarisation due to uncertainty on $\vec\alpha$ is roughly $2.3\%$.
The averaged values correspond to the calibration parameters, $G_0$, $G_1$, and $G_2$, discussed in Ref.~\cite{Wei:2022kem},
up to a convention on the overall rotation, $R_{W}$, discussed above.
Equation~\eqref{eq:alpha.prime} suggests the translation between these calibration parameters and the averaged polarimeter components:
$G_0 = \overline{\alpha}_y$, $G_1 = -\overline{\alpha}_x$, and $G_2 = -\overline{\alpha}_z$.

The polarimeter vector for the charge-conjugated mode is obtained using exact \CP symmetry (see Appendix~\ref{sec:CP} for the details),
\begin{align} \label{eq:Lcbar.alpha.componentes}
    \alpha_x^{\Lcbar} &= -\alpha_x^{\Lc}\,, &
    \alpha_y^{\Lcbar} &=  \alpha_y^{\Lc}\,, &
    \alpha_z^{\Lcbar} &= -\alpha_z^{\Lc}\,.
\end{align}
Any deviation from these relations would indicate \CP violation.
Given that 
\CP violation effects are expected to be negligible for polarisation studies,
the expression for the $\vec\alpha_{\Lcbar}$ obtained under exact \CP symmetry
can be used for the $\Lcbar$ decays (see Appendix~\ref{sec:example} for an example).

\section{Summary} \label{sec:conclusion}

The distribution of the spin state of a fermion contains important information on its production process.
Polarisation can be considered as a tool for enriching the list of observables sensitive to fundamental physics, ranging from the search of new phenomena beyond the Standard Model to studies of the fundamentals of QCD.
To access the polarisation of a fermion in a decay, the transition amplitude needs to be known.
More precisely, one needs to characterise how the kinematic distributions of the decay products reflect the initial polarisation.

In this paper, a model-agnostic representation of the fermion decay rate on the entire space of kinematic dimensions for a multibody decay of a fermion is introduced.
The polarised decay rate has been expressed in terms of Euler angles of the decay-plane orientation and a kinematics-dependent, aligned polarimeter vector.
The distribution of the polarimeter vector has been explored for the dominant $\Lc\to p \Km\pip$ hadronic decay.
Using the recent amplitude analysis results of Ref.~\cite{LHCb-PAPER-2022-002} by \lhcb, the polarimeter vector distribution has been evaluated.
The statistical and model uncertainties are carefully propagated to the inferred quantities based on the original \lhcb analysis.
Knowledge of the polarimeter distribution of the $\Lc$ baryon in its main hadronic decay mode opens many new avenues for testing the Standard Model and determining the properties of strong-interacting systems.
The numerical results provided in this paper facilitate the use of the $\Lc \to \proton \Km \pip$ transition in any practical case without the need to implement amplitude models.

%% file: acknowledgements.tex
\section*{Acknowledgements}
%
%
\noindent We express our gratitude to our colleagues in the CERN
accelerator departments for the excellent performance of the LHC. We
thank the technical and administrative staff at the LHCb
institutes.
We acknowledge support from CERN and from the national agencies:
CAPES, CNPq, FAPERJ and FINEP (Brazil); 
MOST and NSFC (China); 
CNRS/IN2P3 (France); 
BMBF, DFG and MPG (Germany); 
INFN (Italy); 
NWO (Netherlands); 
MNiSW and NCN (Poland); 
MCID/IFA (Romania); 
MICINN (Spain); 
SNSF and SER (Switzerland); 
NASU (Ukraine); 
STFC (United Kingdom); 
DOE NP and NSF (USA).
We acknowledge the computing resources that are provided by CERN, IN2P3
(France), KIT and DESY (Germany), INFN (Italy), SURF (Netherlands),
PIC (Spain), GridPP (United Kingdom), 
CSCS (Switzerland), IFIN-HH (Romania), CBPF (Brazil),
and Polish WLCG (Poland).
We are indebted to the communities behind the multiple open-source
software packages on which we depend.
Individual groups or members have received support from
ARC and ARDC (Australia);
Key Research Program of Frontier Sciences of CAS, CAS PIFI, CAS CCEPP, 
Fundamental Research Funds for the Central Universities, 
and Sci. \& Tech. Program of Guangzhou (China);
Minciencias (Colombia);
EPLANET, Marie Sk\l{}odowska-Curie Actions, ERC and NextGenerationEU (European Union);
A*MIDEX, ANR, IPhU and Labex P2IO, and R\'{e}gion Auvergne-Rh\^{o}ne-Alpes (France);
AvH Foundation (Germany);
ICSC (Italy); 
GVA, XuntaGal, GENCAT, Inditex, InTalent and Prog.~Atracci\'on Talento, CM (Spain);
SRC (Sweden);
the Leverhulme Trust, the Royal Society
 and UKRI (United Kingdom).

%% file: supplementary.tex
\appendix


\section{Decay amplitude for the \texorpdfstring{$\Lc \to p \Km \pip$}{Λc → pπK} decay in the helicity formalism}
\label{sec:DPD}

The helicity amplitude for the $\Lc \to \proton \Km \pip$ decay reads as a sum of three partial wave series, incorporating $K^{**0}$, $\Deltares^{**++}$, and $\Lambdares^{**}$ resonances. Numbering the particles as $\Lc(0) \to \proton(1) \pip(2) \Km(3)$, the aligned amplitude reads
\begin{align} \label{eq:A.DPD.total}
    \mathcal{A}_{\nu,\lambda}(m_{K\pi},m_{pK}) &=
    \sum_{k=1}^3
    \sum_{\nu',\lambda'} 
    d_{\nu,\nu'}^{1/2}(\zeta_{k(1)}^0)\,\mathcal{A}_{\nu',\lambda'}^{\xi_k}\,d_{\lambda',\lambda}^{1/2}(\zeta_{k(1)}^1)\,,
\end{align}
where $d_{\nu,\nu'}^{1/2}$ is the Wigner~$d$-function,
$\zeta^{1}_{k(1)}$ and $\zeta^{0}_{k(1)}$ are the Wigner rotation angles for the proton (particle $1$) and the $\Lc$ baryon (particle 0). The index $\xi_k$ labels the amplitudes for the subsystems, with $k=1$, $k=2$, and $k=3$ representing the subsystem of the $K^{**0}$, $\Deltares^{**++}$, and $\Lambdares^{**}$ resonances, respectively. These amplitudes read
\begin{align} \label{eq:A.DPD.isobar}
    \mathcal{A}^{\xi_k}_{\nu,\lambda} &= \sum_{j,\tau} \delta_{\nu,\tau - \lambda_k}\mathcal{H}^{\Lc \to \xi_k,P_k}_{\tau,\lambda_k} (-1)^{j_k - \lambda_k} \,d^{j}_{\lambda,0} (\theta_{ij}) \, \mathcal{H}^{\xi \to P_i,P_j}_{\lambda_i,\lambda_j}  (-1)^{j_j-\lambda_j}\,.
\end{align}
The helicity couplings in the strong decay of subchannel resonances
obey simple properties with respect to the parity transformation, which reduces the number of couplings in the strong decay of isobars. Moreover, the magnitude of the couplings cannot be determined separately and is set to 1, so that
\begin{align} \label{eq:decay.parity.relation}
    \mathcal{H}^{\Lambdares^{**} \to K p}_{0,1/2} &= 1\,, &
    \mathcal{H}^{\Deltares^{**} \to p\pi}_{1/2,0} &= 1\,,&
    \mathcal{H}^{K^{**} \to \pi K}_{0,0} &= 1, \nonumber \\
    \mathcal{H}^{\Lambdares^{**} \to K p}_{0,-1/2} &= -P_\Lambda (-1)^{j-1/2}\,, &
    \mathcal{H}^{\Deltares^{**} \to p\pi}_{-1/2,0} &= -P_\Delta (-1)^{j-1/2}\,.    &&
\end{align}
Angles with repeated lower indices are zero.
The other angles are computed from the Mandelstam variables using Appendix A of Ref.~\cite{JPAC:2019ufm}.
Note that the angles $\zeta^0$ here are denoted by $\hat{\theta}$ in Ref.~\cite{JPAC:2019ufm}.


\section{Matching to the LHCb model}
\label{sec:LHCb.mapping}

The decay plane used for formulating the amplitude model in Ref.~\cite{LHCb-PAPER-2022-002} is defined by the momenta of the proton and kaon in the $\Lc$ rest frame.
The $z$~axis is set by the proton momentum, while the $y$~axis is orthogonal to the kaon and proton momenta.
The transition amplitude from Ref.~\cite{LHCb-PAPER-2022-002} reads
\begin{align} \label{eq:total.LHCb}
    \mathcal{B}_{\nu,\lambda_p}  &= 
    \sum_{\lambda} \mathcal{K}^{\Lc \to K^{**} p}_{\lambda,\lambda_p} \delta_{\nu,\lambda + \lambda_p}\,d^{J_{K}}_{\lambda,0} (\theta_{K}) \nonumber \\
     &\quad + \sum_{\lambda,\mu_p} \mathcal{K}^{\Lc \to \Lambdares^{**} \pi}_{\lambda} \mathcal{K}^{\Lambda \to p K}_{\mu_p} d^{1/2}_{\nu,\lambda}( \theta_{\Lambda})\, d^{J_{\Lambda}}_{\lambda,\mu_p} (\theta_p^{\Lambda})\, d^{1/2}_{\mu_p,\lambda_p}(-(\theta_p^{\Lambda} + \theta_{\Lambda} + \alpha^{\Lambda}_W)) \nonumber \\
    &\quad + \sum_{\lambda,\mu_p} \mathcal{K}^{\Lc \to \Deltares^{**} K}_{\lambda} \mathcal{K}^{\Delta \to p \pi}_{\mu_p} d^{1/2}_{\nu,\lambda}(\theta_\Delta)\, d^{J_\Delta}_{\lambda,\mu_p} (\theta_p^\Delta)\, d^{1/2}_{\mu_p,\lambda_p}(-(\theta_p^{\Delta} + \theta_\Delta - \alpha^\Delta_W))\,,
\end{align}
where $\mathcal{K}^{A \to BC}_{\lambda_B,\lambda_C}$ represents the helicity coupling constant for the two-body decay $A \to BC$ with $\lambda_B$ and $\lambda_c$ being the helicities of particles $B$ and $C$, respectively. The helicity index for the pseudoscalar particles is omitted.
An algorithm to compute the helicity angles, denoted by the letter $\theta$, can be found in Ref.~\cite{LHCb-PAPER-2022-002}.
The spin-alignment rotation is expressed via the Thomas angle (see Ref.~\cite{Gourgoulhon:2013gua}) denoted by $\alpha_W^\Lambda$ and $\alpha_W^\Delta$.
They are related to the positive angles in Eq.~\eqref{eq:A.DPD.total} and Eq.~\eqref{eq:A.DPD.isobar} as follows:
\begin{align} 
    &&\bar{\theta}_K &= \theta_{23}\,, \nonumber \\
    \theta_{\Lambda} &= \pi - \zeta^0_{1(2)}\,,&\theta_p^{\Lambda} &= -(\pi - \theta_{31})\,, &
    \theta_p^{\Lambda} + \theta_{\Lambda} + \alpha^{\Lambda}_W &= - \zeta_{2(1)}^1\,, \nonumber \\
    \theta_{\Delta} &= -(\pi-\zeta^0_{3(1)})\,,&\theta_p^{\Delta} &= \theta_{12}\,, &
    \theta_p^{\Delta} + \theta_\Delta - \alpha^\Delta_W  &= \zeta_{1(3)}^1\,.
\end{align}

Equation~\eqref{eq:total.LHCb} can be simplified using the properties of the Wigner $d$-functions.
The relation of the full amplitudes from Eq.~\eqref{eq:total.LHCb} and Eq.~\eqref{eq:A.DPD.total} is obtained
by isolating the phase factor that depends on the external helicity indices, which gives
\begin{align}
    \mathcal{B}_{\nu,\lambda_p} &= (-1)^{1/2+\lambda_p} \mathcal{A}_{\nu,-\lambda_p}\,.
\end{align}
The remaining phase factors redefine the product of the helicity couplings, so that the helicity couplings $\mathcal{K}$ from Ref.~\cite{LHCb-PAPER-2022-002} are related to the helicity couplings $\mathcal{H}$ from this paper as
\begin{align} \label{eq:KK=HH}
    \mathcal{K}^{\Lc \to K^{**} p}_{\lambda \lambda_p} &= 
    \mathcal{H}^{\Lc \to K^{**} p}_{\lambda -\lambda_p} (-1)^{J_K}\,, \nonumber \\
    \mathcal{K}^{\Lc \to \Lambdares^{**} \pi}_{-\lambda} \mathcal{K}^{\Lambda \to p K}_{-\mu_p} &=
    \mathcal{H}^{\Lc \to \Lambdares^{**} \pi}_{\lambda} (-1)^{J_\Lambda-1/2} \mathcal{H}^{\Lambda \to p K}_{\mu_p}\,, \nonumber \\
    \mathcal{K}^{\Lc \to \Deltares^{**} K}_{-\lambda} \mathcal{K}^{\Delta \to p \pi}_{-\mu_p} &=
    \mathcal{H}^{\Lc \to \Deltares^{**} K}_{\lambda} \mathcal{H}^{\Delta \to \pi p}_{\mu_p}\,.
\end{align}
The decay couplings are fixed by the parity relation in Eq.~\eqref{eq:decay.parity.relation} and can be eliminated from Eq.~\eqref{eq:KK=HH} leading to the final mapping of the model parameters,
\begin{align}
    \mathcal{H}^{\Lc \to K^{**} p}_{\lambda, \lambda_p} &=
    (-1)^{J_K}\mathcal{K}^{\Lc \to K^{**} p}_{\lambda, -\lambda_p}\,, \nonumber \\
    \mathcal{H}^{\Lc \to \Lambdares^{**} \pi}_{\lambda} &=
    -P_\Lambda \mathcal{K}^{\Lc \to \Lambdares^{**} \pi}_{-\lambda}\,, \nonumber \\
    \mathcal{H}^{\Lc \to \Deltares^{**} K}_{\lambda} &= -P_\Delta(-1)^{J_\Delta-1/2}
    \mathcal{K}^{\Lc \to \Deltares^{**} K}_{-\lambda} \,.
\end{align}

\section{Charge conjugation} \label{sec:CP}

The transition amplitude for the $\Lcbar \to \bar{p} \Kp \pim$ decay is constructed by applying the $C$ and $P$ operators to
Eq.~\eqref{eq:T.gen.DPD}.
The particles are replaced by anti-particles (charge conjugation), the three momenta of the final-state particles are flipped, 
and the helicity values are flipped (parity inversion in the \Lc baryon rest frame).
The amplitude reads
\begin{align}
T_{\nu_0,\lambda}^{\Lcbar}(\vec{p}_1, \vec{p}_2, \vec{p}_3) =
T_{\nu_0,-\lambda}^{\Lc}(-\vec{p}_1, -\vec{p}_2, -\vec{p}_3),
\end{align}
where $\nu_0$ is a spin projection of the \Lc baryon onto the $z$~axis of the production plane. The amplitude is not flipped under parity inversion (see Eq.~3.9 in Ref.~\cite{Chung:1971ri}).

When the momenta of the final-state particles are flipped in space, the polarisation angles transform as
\begin{align}
\phi &\to \left\{
\begin{array}{cl}
\pi+\phi, &\text{ for } \phi < 0,\\
-\pi+\phi, &\text{ for } \phi > 0
\end{array}
\right.
&
\theta&\to \pi-\theta &
\chi &\to 
\left\{
\begin{array}{cl}
-\pi-\chi, &\text{ for } \chi < 0\,,\\
 \pi-\chi, &\text{ for } \chi > 0\,.
\end{array}
\right.
\end{align}
Using properties of the Wigner D-function,
\begin{align}
    D_{\nu_0,\nu}^{1/2}(\phi,\theta,\chi) &\to D_{\nu_0,-\nu}^{1/2}(\phi,\theta,\chi)\,e^{i\nu_0\pi} (-1)^{1/2+\nu_0} e^{i\nu \pi},
\end{align}
we find that the amplitude transforms as
\begin{align}
O^{\nu}_{\lambda} \to e^{-i\nu \pi} O^{-\nu}_{-\lambda}\,.
\end{align}
With the definition of $\vec\alpha$ in Eq.~\eqref{eq:alpha.def},
one finds a sign flip of $\alpha_x$  and $\alpha_z$, while $\alpha_y$ remains the same for both charge modes,
therefore,
\begin{equation}
    \vec\alpha^{\Lcbar} =  R_y(\pi) \vec\alpha^{\Lc}\,,
\end{equation}
which matches Eq.~\eqref{eq:Lcbar.alpha.componentes}.


\section{Example of extending amplitude models with the polarimeter field} \label{sec:example}

Knowledge of the $\Lc\to p\Km \pip$ transition can be incorporated to any decay process involving
the decay as a subprocess using the polarimeter vector.
The $\Bp \to \Lc \Lcbar \Kp$ decay is used as an example to demonstrate how to incorporate the polarimetry fields into the decay amplitude.
The decay amplitude is a function of 12~variables: two Mandelstam invariant variables for the $B$ meson decay, three angles and two Dalitz plot variables for each of the $\Lc$ and \Lcbar decays,
\begin{align} \label{eq:B2LcLcK.A}
&\mathcal{A}_{\lambda,\bar{\lambda}} (m_{\Lc \Kp}^2, m_{\Lcbar \Kp}^2; \phi,\theta,\chi,m_{p\Km}^2,m_{\Km\pip}^2; \bar{\phi},\bar{\theta},\bar{\chi},m_{\bar{p}\Kp}^2,m_{\Kp\pim}^2) = \nonumber \\
    &\quad\sum_{\nu_0,\bar{\nu}_0,\nu,\bar{\nu}} O_{\nu_0,\bar{\nu}_0}^B(m_{\Lc \Kp}^2, m_{\Lcbar \Kp}^2) \nonumber \\
    &\qquad\qquad \times D^{1/2}_{\nu_0,\nu}(\phi,\theta,\chi) O_{\nu,\lambda}^\Lc(m_{p\Km}^2,m_{\Km\pip}^2) \nonumber \\
    &\qquad\qquad \times D^{1/2}_{\bar{\nu}_0,\bar{\nu}}(\bar{\phi},\bar{\theta},\bar{\chi}) O_{\bar{\nu},\bar{\lambda}}^\Lcbar(m_{\bar{p}\Kp}^2,m_{\Kp\pim}^2)\,,
\end{align}
where $\lambda$ and $\bar{\lambda}$ are the helicity of the proton and anti-proton in the \Lc and \Lcbar rest frames, respectively.

Then, the decay rate reads
\begin{align} \label{eq:B2LcLcK.I}
&I(m_{\Lc \Kp}^2, m_{\Lcbar \Kp}^2; \phi,\theta,\chi,m_{p\Km}^2,m_{\Km\pip}^2; \bar{\phi},\bar{\theta},\bar{\chi},m_{\bar{p}\Kp}^2,m_{\Kp\pim}^2) = \nonumber \\
    &\quad\sum_{\nu_0,\bar{\nu}_0,\nu,\bar{\nu}}
    \sum_{\nu_0',\bar{\nu}_0',\nu',\bar{\nu}'}
    O_{\nu_0',\bar{\nu}_0'}^{B*}(m_{\Lc \Kp}^2, m_{\Lcbar \Kp}^2)
    O_{\nu_0,\bar{\nu}_0}^B(m_{\Lc \Kp}^2, m_{\Lcbar \Kp}^2) \nonumber \\
    &\qquad\qquad \times D^{1/2*}_{\nu_0',\nu'}(\phi,\theta,\chi) D^{1/2}_{\nu_0,\nu}(\phi,\theta,\chi) \mathfrak{X}_{\nu',\nu}^\Lc (m_{p\Km}^2,m_{\Km\pip}^2) \nonumber \\
    &\qquad\qquad \times D^{1/2*}_{\bar{\nu}_0',\bar{\nu}'}(\bar{\phi},\bar{\theta},\bar{\chi}) D^{1/2}_{\bar{\nu}_0,\bar{\nu}}(\bar{\phi},\bar{\theta},\bar{\chi})\,
    \mathfrak{X}_{\bar{\nu}',\bar{\nu}}^\Lcbar(m_{\bar{p}\Kp}^2,m_{\Kp\pim}^2)\,,
\end{align}
where $\mathfrak{X}$ is the matrix given by Eq.~\eqref{eq:polarimetrty.matrix}.
If we integrate over the degrees of freedom in the \Lc and \Lcbar decays, the expression reduces to a simple Dalitz-plot analysis,
\begin{align}  \label{eq:B2LcLcK.I0}
&I^{\text{(just Dalitz)}}(m_{\Lc \Kp}^2, m_{\Lcbar \Kp}^2) = \sum_{\nu_0,\bar{\nu}_0}
    \left| O_{\nu_0,\bar{\nu}_0}^{B}(m_{\Lc \Kp}^2, m_{\Lcbar \Kp}^2) \right|^2\,.
\end{align}

%% file: supplementary-app.tex
\clearpage

\section*{Supplementary material for LHCb-PAPER-2022-044}
\label{sec:Supplementary-App}

Numerical values for $I_0$, $\alpha_x$, $\alpha_y$, and $\alpha_z$ have been computed over a two-dimensional grid
of the Dalitz plot coordinates and are provided in easy-readable JSON files. The grid dimension is chosen to be \mbox{$100\!\times\!100$} values as a compromise between the precision and the size of the output set. This supplemental material can be downloaded at \url{https://lc2pkpi-polarimetry.docs.cern.ch/_static/export/polarimetry-field.tar.gz}.
Each JSON file provides the following data entries:
\begin{itemize}
    \item \textbf{\texttt{metadata}}: information about the amplitude model with which the field and intensity was computed, including \texttt{model description},  \texttt{parameters}, and \texttt{reference subsystem} that was used to align the amplitude models with Dalitz-plot decomposition.
    \item \textbf{\texttt{m\^{}2\_Kpi}}: an array of 100 values for $\sigma_1 = m^2(K^- \pi^+)$ that span the $x$~axis of the Dalitz grid.
    \item \textbf{\texttt{m\^{}2\_pK}}: an array of 100 values for $\sigma_2 = m^2(p K^-)$ that span the $y$~axis of the Dalitz grid.
    \item \textbf{\texttt{alpha\_x}}, \textbf{\texttt{alpha\_y}}, \textbf{\texttt{alpha\_z}}: computed values for $\alpha_x$, $\alpha_y$, $\alpha_z$ over the \mbox{$100\!\times\!100$} grid array.
    \item \textbf{\texttt{intensity}}: computed unpolarised intensity $I_0$ over the \mbox{$100\!\times\!100$} grid array.
\end{itemize}
Values on the grid that lie outside the phase space are given as \texttt{NaN}.

\clearpage

%% file: references.bbl
\ifx\mcitethebibliography\mciteundefinedmacro
\PackageError{LHCb.bst}{mciteplus.sty has not been loaded}
{This bibstyle requires the use of the mciteplus package.}\fi
\providecommand{\href}[2]{#2}

%% file: Authorship_LHCb-PAPER-2022-044.tex
\centerline
{\large\bf LHCb collaboration}
\begin
{flushleft}
\small
R.~Aaij$^{32}$\lhcborcid{0000-0003-0533-1952},
A.S.W.~Abdelmotteleb$^{50}$\lhcborcid{0000-0001-7905-0542},
C.~Abellan~Beteta$^{44}$,
F.~Abudin{\'e}n$^{50}$\lhcborcid{0000-0002-6737-3528},
T.~Ackernley$^{54}$\lhcborcid{0000-0002-5951-3498},
B.~Adeva$^{40}$\lhcborcid{0000-0001-9756-3712},
M.~Adinolfi$^{48}$\lhcborcid{0000-0002-1326-1264},
P.~Adlarson$^{77}$\lhcborcid{0000-0001-6280-3851},
H.~Afsharnia$^{9}$,
C.~Agapopoulou$^{13}$\lhcborcid{0000-0002-2368-0147}
% ...

{\footnotesize \it

% ...
$^{9}$Universit{\'e} Clermont Auvergne, CNRS/IN2P3, LPC, Clermont-Ferrand, France\\
% ...
$^{13}$LPNHE, Sorbonne Universit{\'e}, Paris Diderot Sorbonne Paris Cit{\'e}, CNRS/IN2P3, Paris, France\\
% ...
$^{32}$Nikhef National Institute for Subatomic Physics, Amsterdam, Netherlands\\
% ...
$^{40}$Instituto Galego de F{\'\i}sica de Altas Enerx{\'\i}as (IGFAE), Universidade de Santiago de Compostela, Santiago de Compostela, Spain\\
% ...
$^{44}$Physik-Institut, Universit{\"a}t Z{\"u}rich, Z{\"u}rich, Switzerland\\
% ...
$^{48}$H.H. Wills Physics Laboratory, University of Bristol, Bristol, United Kingdom\\
% ...
$^{50}$Department of Physics, University of Warwick, Coventry, United Kingdom\\
% ...
$^{53}$School of Physics and Astronomy, University of Glasgow, Glasgow, United Kingdom\\
$^{54}$Oliver Lodge Laboratory, University of Liverpool, Liverpool, United Kingdom\\
% ...
$^{77}$Department of Physics and Astronomy, Uppsala University, Uppsala, Sweden, associated to $^{53}$\\
% ...
}
\end{flushleft}

%% file: main.bbl
\begin{mcitethebibliography}{10}
\mciteSetBstSublistMode{n}
\mciteSetBstMaxWidthForm{subitem}{\alph{mcitesubitemcount})}
\mciteSetBstSublistLabelBeginEnd{\mcitemaxwidthsubitemform\space}
{\relax}{\relax}

\bibitem{LHCb-PAPER-2015-029}
LHCb collaboration, R.~Aaij {\em et~al.},
  \ifthenelse{\boolean{articletitles}}{\emph{{Observation of $\jpsi\proton$
  resonances consistent with pentaquark states in
  \mbox{\decay{\Lb}{\jpsi\proton\Km}} decays}},
  }{}\href{https://doi.org/10.1103/PhysRevLett.115.072001}{Phys.\ Rev.\ Lett.\
  \textbf{115} (2015) 072001},
  \href{http://arxiv.org/abs/1507.03414}{{\normalfont\ttfamily
  arXiv:1507.03414}}\relax
\mciteBstWouldAddEndPuncttrue
\mciteSetBstMidEndSepPunct{\mcitedefaultmidpunct}
{\mcitedefaultendpunct}{\mcitedefaultseppunct}\relax
\EndOfBibitem
\bibitem{LHCb-PAPER-2021-012}
LHCb collaboration, R.~Aaij {\em et~al.},
  \ifthenelse{\boolean{articletitles}}{\emph{{Observation of excited $\Omegac$
  baryons in \mbox{$\Omegab \to \Xic^+ \Km \pip$} decays}},
  }{}\href{https://doi.org/10.1103/PhysRevD.104.L091102}{Phys.\ Rev.\
  \textbf{D104} (2021) L091102},
  \href{http://arxiv.org/abs/2107.03419}{{\normalfont\ttfamily
  arXiv:2107.03419}}\relax
\mciteBstWouldAddEndPuncttrue
\mciteSetBstMidEndSepPunct{\mcitedefaultmidpunct}
{\mcitedefaultendpunct}{\mcitedefaultseppunct}\relax
\EndOfBibitem
\bibitem{Brambilla:2010cs}
N.~Brambilla {\em et~al.}, \ifthenelse{\boolean{articletitles}}{\emph{{Heavy
  Quarkonium: progress, puzzles, and opportunities}},
  }{}\href{https://doi.org/10.1140/epjc/s10052-010-1534-9}{Eur.\ Phys.\ J.\
  \textbf{C71} (2011) 1534},
  \href{http://arxiv.org/abs/1010.5827}{{\normalfont\ttfamily
  arXiv:1010.5827}}\relax
\mciteBstWouldAddEndPuncttrue
\mciteSetBstMidEndSepPunct{\mcitedefaultmidpunct}
{\mcitedefaultendpunct}{\mcitedefaultseppunct}\relax
\EndOfBibitem
\bibitem{Faccioli:2010kd}
P.~Faccioli, C.~Lourenco, J.~Seixas, and H.~K. Wohri,
  \ifthenelse{\boolean{articletitles}}{\emph{{Towards the experimental
  clarification of quarkonium polarization}},
  }{}\href{https://doi.org/10.1140/epjc/s10052-010-1420-5}{Eur.\ Phys.\ J.\
  \textbf{C69} (2010) 657},
  \href{http://arxiv.org/abs/1006.2738}{{\normalfont\ttfamily
  arXiv:1006.2738}}\relax
\mciteBstWouldAddEndPuncttrue
\mciteSetBstMidEndSepPunct{\mcitedefaultmidpunct}
{\mcitedefaultendpunct}{\mcitedefaultseppunct}\relax
\EndOfBibitem
\bibitem{Butenschoen:2012px}
M.~Butenschoen and B.~A. Kniehl,
  \ifthenelse{\boolean{articletitles}}{\emph{{$J/\psi$ polarization at Tevatron
  and LHC: nonrelativistic-QCD factorization at the crossroads}},
  }{}\href{https://doi.org/10.1103/PhysRevLett.108.172002}{Phys.\ Rev.\ Lett.\
  \textbf{108} (2012) 172002},
  \href{http://arxiv.org/abs/1201.1872}{{\normalfont\ttfamily
  arXiv:1201.1872}}\relax
\mciteBstWouldAddEndPuncttrue
\mciteSetBstMidEndSepPunct{\mcitedefaultmidpunct}
{\mcitedefaultendpunct}{\mcitedefaultseppunct}\relax
\EndOfBibitem
\bibitem{Konig:1993wz}
B.~Konig, J.~G. Korner, and M.~Kramer,
  \ifthenelse{\boolean{articletitles}}{\emph{{On the determination of the $b
  \to c$ handedness using nonleptonic $\varLambda_c$ decays}},
  }{}\href{https://doi.org/10.1103/PhysRevD.49.2363}{Phys.\ Rev.\  \textbf{D49}
  (1994) 2363},
  \href{http://arxiv.org/abs/hep-ph/9310263}{{\normalfont\ttfamily
  arXiv:hep-ph/9310263}}\relax
\mciteBstWouldAddEndPuncttrue
\mciteSetBstMidEndSepPunct{\mcitedefaultmidpunct}
{\mcitedefaultendpunct}{\mcitedefaultseppunct}\relax
\EndOfBibitem
\bibitem{Dutta:2015ueb}
R.~Dutta, \ifthenelse{\boolean{articletitles}}{\emph{{$\varLambda_b \to
  (\varLambda_c,\,p)\,\tau\,\nu$ decays within standard model and beyond}},
  }{}\href{https://doi.org/10.1103/PhysRevD.93.054003}{Phys.\ Rev.\
  \textbf{D93} (2016) 054003},
  \href{http://arxiv.org/abs/1512.04034}{{\normalfont\ttfamily
  arXiv:1512.04034}}\relax
\mciteBstWouldAddEndPuncttrue
\mciteSetBstMidEndSepPunct{\mcitedefaultmidpunct}
{\mcitedefaultendpunct}{\mcitedefaultseppunct}\relax
\EndOfBibitem
\bibitem{Shivashankara:2015cta}
S.~Shivashankara, W.~Wu, and A.~Datta,
  \ifthenelse{\boolean{articletitles}}{\emph{{$\varLambda_b \to \varLambda_c
  \tau \bar{\nu}_{\tau}$ decay in the standard model and with new physics}},
  }{}\href{https://doi.org/10.1103/PhysRevD.91.115003}{Phys.\ Rev.\
  \textbf{D91} (2015) 115003},
  \href{http://arxiv.org/abs/1502.07230}{{\normalfont\ttfamily
  arXiv:1502.07230}}\relax
\mciteBstWouldAddEndPuncttrue
\mciteSetBstMidEndSepPunct{\mcitedefaultmidpunct}
{\mcitedefaultendpunct}{\mcitedefaultseppunct}\relax
\EndOfBibitem
\bibitem{Li:2016pdv}
X.-Q. Li, Y.-D. Yang, and X.~Zhang,
  \ifthenelse{\boolean{articletitles}}{\emph{{$\varLambda_b\to \varLambda_c\tau
  \bar{\nu}_\tau $ decay in scalar and vector leptoquark scenarios}},
  }{}\href{https://doi.org/10.1007/JHEP02(2017)068}{JHEP \textbf{02} (2017)
  068}, \href{http://arxiv.org/abs/1611.01635}{{\normalfont\ttfamily
  arXiv:1611.01635}}\relax
\mciteBstWouldAddEndPuncttrue
\mciteSetBstMidEndSepPunct{\mcitedefaultmidpunct}
{\mcitedefaultendpunct}{\mcitedefaultseppunct}\relax
\EndOfBibitem
\bibitem{Datta:2017aue}
A.~Datta, S.~Kamali, S.~Meinel, and A.~Rashed,
  \ifthenelse{\boolean{articletitles}}{\emph{{Phenomenology of $
  {\varLambda}_b\to {\varLambda}_c\tau {\overline{\nu}}_{\tau } $ using lattice
  QCD calculations}}, }{}\href{https://doi.org/10.1007/JHEP08(2017)131}{JHEP
  \textbf{08} (2017) 131},
  \href{http://arxiv.org/abs/1702.02243}{{\normalfont\ttfamily
  arXiv:1702.02243}}\relax
\mciteBstWouldAddEndPuncttrue
\mciteSetBstMidEndSepPunct{\mcitedefaultmidpunct}
{\mcitedefaultendpunct}{\mcitedefaultseppunct}\relax
\EndOfBibitem
\bibitem{Ray:2018hrx}
A.~Ray, S.~Sahoo, and R.~Mohanta,
  \ifthenelse{\boolean{articletitles}}{\emph{{Probing new physics in
  semileptonic $\varLambda_b$ decays}},
  }{}\href{https://doi.org/10.1103/PhysRevD.99.015015}{Phys.\ Rev.\
  \textbf{D99} (2019) 015015},
  \href{http://arxiv.org/abs/1812.08314}{{\normalfont\ttfamily
  arXiv:1812.08314}}\relax
\mciteBstWouldAddEndPuncttrue
\mciteSetBstMidEndSepPunct{\mcitedefaultmidpunct}
{\mcitedefaultendpunct}{\mcitedefaultseppunct}\relax
\EndOfBibitem
\bibitem{DiSalvo:2018ngq}
E.~Di~Salvo, F.~Fontanelli, and Z.~J. Ajaltouni,
  \ifthenelse{\boolean{articletitles}}{\emph{{Detailed study of the
  $\varLambda_b \to \varLambda_c \tau {\bar \nu}_{\tau}$} decay},
  }{}\href{https://doi.org/10.1142/S0217751X18501695}{Int.\ J.\ Mod.\ Phys.\
  \textbf{A33} (2018) 1850169},
  \href{http://arxiv.org/abs/1804.05592}{{\normalfont\ttfamily
  arXiv:1804.05592}}\relax
\mciteBstWouldAddEndPuncttrue
\mciteSetBstMidEndSepPunct{\mcitedefaultmidpunct}
{\mcitedefaultendpunct}{\mcitedefaultseppunct}\relax
\EndOfBibitem
\bibitem{Penalva:2019rgt}
N.~Penalva, E.~Hern\'andez, and J.~Nieves,
  \ifthenelse{\boolean{articletitles}}{\emph{{Further tests of lepton flavour
  universality from the charged lepton energy distribution in $b\to c$
  semileptonic decays: The case of $\varLambda_b\to \varLambda_c \ell
  \bar\nu_\ell$}}, }{}\href{https://doi.org/10.1103/PhysRevD.100.113007}{Phys.\
  Rev.\  \textbf{D100} (2019) 113007},
  \href{http://arxiv.org/abs/1908.02328}{{\normalfont\ttfamily
  arXiv:1908.02328}}\relax
\mciteBstWouldAddEndPuncttrue
\mciteSetBstMidEndSepPunct{\mcitedefaultmidpunct}
{\mcitedefaultendpunct}{\mcitedefaultseppunct}\relax
\EndOfBibitem
\bibitem{Ferrillo:2019owd}
M.~Ferrillo, A.~Mathad, P.~Owen, and N.~Serra,
  \ifthenelse{\boolean{articletitles}}{\emph{{Probing effects of new physics in
  $\varLambda^0_{b}\to\varLambda^+_{c}\mu^{-}\bar{\nu}_{\mu}$ decays}},
  }{}\href{https://doi.org/10.1007/JHEP12(2019)148}{JHEP \textbf{12} (2019)
  148}, \href{http://arxiv.org/abs/1909.04608}{{\normalfont\ttfamily
  arXiv:1909.04608}}\relax
\mciteBstWouldAddEndPuncttrue
\mciteSetBstMidEndSepPunct{\mcitedefaultmidpunct}
{\mcitedefaultendpunct}{\mcitedefaultseppunct}\relax
\EndOfBibitem
\bibitem{LHCb-PAPER-2022-002}
LHCb collaboration, R.~Aaij {\em et~al.},
  \ifthenelse{\boolean{articletitles}}{\emph{{Amplitude analysis of the $\Lc
  \to \proton \Km\pip$ decay and $\Lc$ baryon polarization measurement in
  semileptonic beauty hadron decays}},
  }{}\href{https://doi.org/10.1103/PhysRevD.108.012023}{Phys.\ Rev.\
  \textbf{D108} (2023) 012023},
  \href{http://arxiv.org/abs/2208.03262}{{\normalfont\ttfamily
  arXiv:2208.03262}}\relax
\mciteBstWouldAddEndPuncttrue
\mciteSetBstMidEndSepPunct{\mcitedefaultmidpunct}
{\mcitedefaultendpunct}{\mcitedefaultseppunct}\relax
\EndOfBibitem
\bibitem{Baryshevsky:2016cul}
V.~G. Baryshevsky, \ifthenelse{\boolean{articletitles}}{\emph{{The possibility
  to measure the magnetic moments of short-lived particles (charm and beauty
  baryons) at LHC and FCC energies using the phenomenon of spin rotation in
  crystals}}, }{}\href{https://doi.org/10.1016/j.physletb.2016.04.025}{Phys.\
  Lett.\  \textbf{B757} (2016) 426}\relax
\mciteBstWouldAddEndPuncttrue
\mciteSetBstMidEndSepPunct{\mcitedefaultmidpunct}
{\mcitedefaultendpunct}{\mcitedefaultseppunct}\relax
\EndOfBibitem
\bibitem{Botella:2016ksl}
F.~J. Botella {\em et~al.}, \ifthenelse{\boolean{articletitles}}{\emph{{On the
  search for the electric dipole moment of strange and charm baryons at LHC}},
  }{}\href{https://doi.org/10.1140/epjc/s10052-017-4679-y}{Eur.\ Phys.\ J.\
  \textbf{C77} (2017) 181},
  \href{http://arxiv.org/abs/1612.06769}{{\normalfont\ttfamily
  arXiv:1612.06769}}\relax
\mciteBstWouldAddEndPuncttrue
\mciteSetBstMidEndSepPunct{\mcitedefaultmidpunct}
{\mcitedefaultendpunct}{\mcitedefaultseppunct}\relax
\EndOfBibitem
\bibitem{Fomin:2017ltw}
A.~S. Fomin {\em et~al.},
  \ifthenelse{\boolean{articletitles}}{\emph{{Feasibility of measuring the
  magnetic dipole moments of the charm baryons at the LHC using bent
  crystals}}, }{}\href{https://doi.org/10.1007/JHEP08(2017)120}{JHEP
  \textbf{08} (2017) 120},
  \href{http://arxiv.org/abs/1705.03382}{{\normalfont\ttfamily
  arXiv:1705.03382}}\relax
\mciteBstWouldAddEndPuncttrue
\mciteSetBstMidEndSepPunct{\mcitedefaultmidpunct}
{\mcitedefaultendpunct}{\mcitedefaultseppunct}\relax
\EndOfBibitem
\bibitem{Fomin:2019wuw}
A.~S. Fomin {\em et~al.}, \ifthenelse{\boolean{articletitles}}{\emph{{The
  prospect of charm quark magnetic moment determination}},
  }{}\href{https://doi.org/10.1140/epjc/s10052-020-7891-0}{Eur.\ Phys.\ J.\ C
  \textbf{80} (2020) 358},
  \href{http://arxiv.org/abs/1909.04654}{{\normalfont\ttfamily
  arXiv:1909.04654}}\relax
\mciteBstWouldAddEndPuncttrue
\mciteSetBstMidEndSepPunct{\mcitedefaultmidpunct}
{\mcitedefaultendpunct}{\mcitedefaultseppunct}\relax
\EndOfBibitem
\bibitem{Aiola:2020yam}
S.~Aiola {\em et~al.}, \ifthenelse{\boolean{articletitles}}{\emph{{Progress
  towards the first measurement of charm baryon dipole moments}},
  }{}\href{https://doi.org/10.1103/PhysRevD.103.072003}{Phys.\ Rev.\
  \textbf{D103} (2021) 072003},
  \href{http://arxiv.org/abs/2010.11902}{{\normalfont\ttfamily
  arXiv:2010.11902}}\relax
\mciteBstWouldAddEndPuncttrue
\mciteSetBstMidEndSepPunct{\mcitedefaultmidpunct}
{\mcitedefaultendpunct}{\mcitedefaultseppunct}\relax
\EndOfBibitem
\bibitem{PDG2022}
Particle Data Group, R.~L. Workman {\em et~al.},
  \ifthenelse{\boolean{articletitles}}{\emph{{\href{http://pdg.lbl.gov/}{Review
  of particle physics}}}, }{}\href{https://doi.org/10.1093/ptep/ptac097}{Prog.\
  Theor.\ Exp.\ Phys.\  \textbf{2022} (2022) 083C01}\relax
\mciteBstWouldAddEndPuncttrue
\mciteSetBstMidEndSepPunct{\mcitedefaultmidpunct}
{\mcitedefaultendpunct}{\mcitedefaultseppunct}\relax
\EndOfBibitem
\bibitem{Tsai:1971vv}
Y.-S. Tsai, \ifthenelse{\boolean{articletitles}}{\emph{{Decay correlations of
  heavy leptons in $\ep \en \to \ellp \ellm$}},
  }{}\href{https://doi.org/10.1103/PhysRevD.4.2821}{Phys.\ Rev.\  \textbf{D4}
  (1971) 2821}, [Erratum: Phys.Rev.D 13, 771 (1976)]\relax
\mciteBstWouldAddEndPuncttrue
\mciteSetBstMidEndSepPunct{\mcitedefaultmidpunct}
{\mcitedefaultendpunct}{\mcitedefaultseppunct}\relax
\EndOfBibitem
\bibitem{Kuhn:1991cc}
J.~H. Kuhn and E.~Mirkes, \ifthenelse{\boolean{articletitles}}{\emph{{Angular
  distributions in semileptonic tau decays}},
  }{}\href{https://doi.org/10.1016/0370-2693(92)91791-7}{Phys.\ Lett.\
  \textbf{B286} (1992) 381}\relax
\mciteBstWouldAddEndPuncttrue
\mciteSetBstMidEndSepPunct{\mcitedefaultmidpunct}
{\mcitedefaultendpunct}{\mcitedefaultseppunct}\relax
\EndOfBibitem
\bibitem{Davier:1992nw}
M.~Davier, L.~Duflot, F.~Le~Diberder, and A.~Rouge,
  \ifthenelse{\boolean{articletitles}}{\emph{{The optimal method for the
  measurement of tau polarization}},
  }{}\href{https://doi.org/10.1016/0370-2693(93)90101-M}{Phys.\ Lett.\
  \textbf{B306} (1993) 411}\relax
\mciteBstWouldAddEndPuncttrue
\mciteSetBstMidEndSepPunct{\mcitedefaultmidpunct}
{\mcitedefaultendpunct}{\mcitedefaultseppunct}\relax
\EndOfBibitem
\bibitem{Kuhn:1995nn}
J.~H. Kuhn, \ifthenelse{\boolean{articletitles}}{\emph{{Tau polarimetry with
  multi - meson states}},
  }{}\href{https://doi.org/10.1103/PhysRevD.52.3128}{Phys.\ Rev.\  \textbf{D52}
  (1995) 3128},
  \href{http://arxiv.org/abs/hep-ph/9505303}{{\normalfont\ttfamily
  arXiv:hep-ph/9505303}}\relax
\mciteBstWouldAddEndPuncttrue
\mciteSetBstMidEndSepPunct{\mcitedefaultmidpunct}
{\mcitedefaultendpunct}{\mcitedefaultseppunct}\relax
\EndOfBibitem
\bibitem{Kuhn:1982di}
J.~H. Kuhn and F.~Wagner,
  \ifthenelse{\boolean{articletitles}}{\emph{{Semileptonic decays of the $\tau$
  lepton}}, }{}\href{https://doi.org/10.1016/0550-3213(84)90522-4}{Nucl.\
  Phys.\  \textbf{B236} (1984) 16}\relax
\mciteBstWouldAddEndPuncttrue
\mciteSetBstMidEndSepPunct{\mcitedefaultmidpunct}
{\mcitedefaultendpunct}{\mcitedefaultseppunct}\relax
\EndOfBibitem
\bibitem{Kuhn:1993ra}
J.~H. Kuhn, \ifthenelse{\boolean{articletitles}}{\emph{{Tau kinematics from
  impact parameters}},
  }{}\href{https://doi.org/10.1016/0370-2693(93)90019-E}{Phys.\ Lett.\
  \textbf{B313} (1993) 458},
  \href{http://arxiv.org/abs/hep-ph/9307269}{{\normalfont\ttfamily
  arXiv:hep-ph/9307269}}\relax
\mciteBstWouldAddEndPuncttrue
\mciteSetBstMidEndSepPunct{\mcitedefaultmidpunct}
{\mcitedefaultendpunct}{\mcitedefaultseppunct}\relax
\EndOfBibitem
\bibitem{Hagiwara:1989fn}
K.~Hagiwara, A.~D. Martin, and D.~Zeppenfeld,
  \ifthenelse{\boolean{articletitles}}{\emph{{$\tau$ polarization measurements
  at LEP and SLC}},
  }{}\href{https://doi.org/10.1016/0370-2693(90)90120-U}{Phys.\ Lett.\
  \textbf{B235} (1990) 198}\relax
\mciteBstWouldAddEndPuncttrue
\mciteSetBstMidEndSepPunct{\mcitedefaultmidpunct}
{\mcitedefaultendpunct}{\mcitedefaultseppunct}\relax
\EndOfBibitem
\bibitem{Jezabek:1992ke}
M.~Jezabek, K.~Rybicki, and R.~Rylko,
  \ifthenelse{\boolean{articletitles}}{\emph{{Experimental study of spin
  effects in hadroproduction and decay of $\varLambda_c^+$}},
  }{}\href{https://doi.org/10.1016/0370-2693(92)90177-6}{Phys.\ Lett.\
  \textbf{B286} (1992) 175}\relax
\mciteBstWouldAddEndPuncttrue
\mciteSetBstMidEndSepPunct{\mcitedefaultmidpunct}
{\mcitedefaultendpunct}{\mcitedefaultseppunct}\relax
\EndOfBibitem
\bibitem{Jezabek:1992vi}
M.~Jezabek, K.~Rybicki, and R.~Rylko,
  \ifthenelse{\boolean{articletitles}}{\emph{{A measurement of $\varLambda_c^+$
  spin using the \mbox{$\varLambda_c^+ \rightarrow p K^- \pi^+$} decay
  channel}}, }{}Acta Phys.\ Polon.\  \textbf{B23} (1992) 771\relax
\mciteBstWouldAddEndPuncttrue
\mciteSetBstMidEndSepPunct{\mcitedefaultmidpunct}
{\mcitedefaultendpunct}{\mcitedefaultseppunct}\relax
\EndOfBibitem
\bibitem{E791:1999ajq}
E791 collaboration, E.~M. Aitala {\em et~al.},
  \ifthenelse{\boolean{articletitles}}{\emph{{Multidimensional resonance
  analysis of \mbox{$\varLambda_c^+ \rightarrow p K^- \pi^+$}}},
  }{}\href{https://doi.org/10.1016/S0370-2693(99)01397-0}{Phys.\ Lett.\
  \textbf{B471} (2000) 449},
  \href{http://arxiv.org/abs/hep-ex/9912003}{{\normalfont\ttfamily
  arXiv:hep-ex/9912003}}\relax
\mciteBstWouldAddEndPuncttrue
\mciteSetBstMidEndSepPunct{\mcitedefaultmidpunct}
{\mcitedefaultendpunct}{\mcitedefaultseppunct}\relax
\EndOfBibitem
\bibitem{Fox:1999ja}
G.~F. Fox, {\em {Multidimensional resonance analysis of $\varLambda^+_c \to p
  K^- \pi^+$}}, PhD thesis, South Carolina U., 1999,
  doi:~\href{https://doi.org/10.2172/1371870}{10.2172/1371870}\relax
\mciteBstWouldAddEndPuncttrue
\mciteSetBstMidEndSepPunct{\mcitedefaultmidpunct}
{\mcitedefaultendpunct}{\mcitedefaultseppunct}\relax
\EndOfBibitem
\bibitem{Bjorken:1988ya}
J.~D. Bjorken, \ifthenelse{\boolean{articletitles}}{\emph{{Spin-dependent
  decays of the $\varLambda_c$}},
  }{}\href{https://doi.org/10.1103/PhysRevD.40.1513}{Phys.\ Rev.\  \textbf{D40}
  (1989) 1513}\relax
\mciteBstWouldAddEndPuncttrue
\mciteSetBstMidEndSepPunct{\mcitedefaultmidpunct}
{\mcitedefaultendpunct}{\mcitedefaultseppunct}\relax
\EndOfBibitem
\bibitem{Wei:2022kem}
D.-H. Wei, Y.-X. Yang, and R.-G. Ping,
  \ifthenelse{\boolean{articletitles}}{\emph{{Using $\Lc \to \proton \Km \pip$
  as a spin polarimeter}},
  }{}\href{https://doi.org/10.1088/1674-1137/ac5e93}{Chin.\ Phys.\
  \textbf{C46} (2022) 074002}\relax
\mciteBstWouldAddEndPuncttrue
\mciteSetBstMidEndSepPunct{\mcitedefaultmidpunct}
{\mcitedefaultendpunct}{\mcitedefaultseppunct}\relax
\EndOfBibitem
\bibitem{JPAC:2019ufm}
JPAC collaboration, M.~Mikhasenko {\em et~al.},
  \ifthenelse{\boolean{articletitles}}{\emph{{Dalitz-plot decomposition for
  three-body decays}},
  }{}\href{https://doi.org/10.1103/PhysRevD.101.034033}{Phys.\ Rev.\
  \textbf{D101} (2020) 034033},
  \href{http://arxiv.org/abs/1910.04566}{{\normalfont\ttfamily
  arXiv:1910.04566}}\relax
\mciteBstWouldAddEndPuncttrue
\mciteSetBstMidEndSepPunct{\mcitedefaultmidpunct}
{\mcitedefaultendpunct}{\mcitedefaultseppunct}\relax
\EndOfBibitem
\bibitem{Cornwell:1997ke}
J.~F. Cornwell, {\em {Group {Theory} in {Physics}: An {Introduction}}},
  \href{https://doi.org/10.1016/B978-0-12-189800-7.X5000-6}{ Academic Press,
  {San Diego, CA}, 1997}\relax
\mciteBstWouldAddEndPuncttrue
\mciteSetBstMidEndSepPunct{\mcitedefaultmidpunct}
{\mcitedefaultendpunct}{\mcitedefaultseppunct}\relax
\EndOfBibitem
\bibitem{Rouge:1991pm}
A.~Rouge, \ifthenelse{\boolean{articletitles}}{\emph{Tau decays as polarization
  analysers}, }{} in {\em Workshop on {{Tau Lepton Physics}} ({{TAU}} 90)},
  213--222, 1991\relax
\mciteBstWouldAddEndPuncttrue
\mciteSetBstMidEndSepPunct{\mcitedefaultmidpunct}
{\mcitedefaultendpunct}{\mcitedefaultseppunct}\relax
\EndOfBibitem
\bibitem{Dalitz:1953cp}
R.~H. Dalitz, \ifthenelse{\boolean{articletitles}}{\emph{{On the analysis of
  \Ptau-meson data and the nature of the \Ptau-meson}},
  }{}\href{https://doi.org/10.1080/14786441008520365}{Phil.\ Mag.\ Ser.\ 7
  \textbf{44} (1953) 1068}\relax
\mciteBstWouldAddEndPuncttrue
\mciteSetBstMidEndSepPunct{\mcitedefaultmidpunct}
{\mcitedefaultendpunct}{\mcitedefaultseppunct}\relax
\EndOfBibitem
\bibitem{Gourgoulhon:2013gua}
E.~Gourgoulhon, {\em {Special Relativity in General Frames}},
  \href{https://doi.org/10.1007/978-3-642-37276-6}{ Graduate Texts in Physics,
  Springer, Berlin, Heidelberg, 2013}\relax
\mciteBstWouldAddEndPuncttrue
\mciteSetBstMidEndSepPunct{\mcitedefaultmidpunct}
{\mcitedefaultendpunct}{\mcitedefaultseppunct}\relax
\EndOfBibitem
\bibitem{Dedu:2742640}
V.~G. Dedu, \ifthenelse{\boolean{articletitles}}{\emph{{Extraction of
  polarization sensitivity in charm-baryon three-body decays in LHCb}},
  }{}\href{https://doi.org/10.17181/CERN-STUDENTS-Note-2020-031}{CERN Document
  Server (2020) }\relax
\mciteBstWouldAddEndPuncttrue
\mciteSetBstMidEndSepPunct{\mcitedefaultmidpunct}
{\mcitedefaultendpunct}{\mcitedefaultseppunct}\relax
\EndOfBibitem
\bibitem{polarimetry.COMPWA:2022xyz}
R.~E. de~Boer, M.~Mikhasenko, and M.~Fritsch,
  \ifthenelse{\boolean{articletitles}}{\emph{{$\Lambda^+_\mathrm{c}$
  polarimetry using the dominant hadronic mode}}, }{} 2023.
\newblock
  doi:~\href{https://doi.org/10.5281/zenodo.7549056}{10.5281/zenodo.7549056}\relax
\mciteBstWouldAddEndPuncttrue
\mciteSetBstMidEndSepPunct{\mcitedefaultmidpunct}
{\mcitedefaultendpunct}{\mcitedefaultseppunct}\relax
\EndOfBibitem
\bibitem{Fritsch:2022compwa}
M.~Fritsch {\em et~al.}, \ifthenelse{\boolean{articletitles}}{\emph{{Common
  Partial Wave Analysis: a collaboration-independent organisation for amplitude
  analysis software}}, }{} 2022.
\newblock
  doi:~\href{https://doi.org/10.5281/zenodo.6908149}{10.5281/zenodo.6908149}\relax
\mciteBstWouldAddEndPuncttrue
\mciteSetBstMidEndSepPunct{\mcitedefaultmidpunct}
{\mcitedefaultendpunct}{\mcitedefaultseppunct}\relax
\EndOfBibitem
\bibitem{Mikhasenko:2022xyz}
M.~Mikhasenko,
  \ifthenelse{\boolean{articletitles}}{\emph{{\texttt{ThreeBodyDecay.jl}}
  {Julia} implementation of the {Dalitz}-plot decomposition}, }{} 2022.
\newblock
  doi:~\href{https://doi.org/10.5281/zenodo.7256812}{10.5281/zenodo.7256812}\relax
\mciteBstWouldAddEndPuncttrue
\mciteSetBstMidEndSepPunct{\mcitedefaultmidpunct}
{\mcitedefaultendpunct}{\mcitedefaultseppunct}\relax
\EndOfBibitem
\bibitem{Chung:1971ri}
S.~U. Chung, {\em {Spin formalisms}},
  \href{https://doi.org/10.5170/CERN-1971-008}{ CERN Academic Training Lecture,
  CERN, Geneva, 1971}.
\newblock CERN, Geneva, 1969 - 1970\relax
\mciteBstWouldAddEndPuncttrue
\mciteSetBstMidEndSepPunct{\mcitedefaultmidpunct}
{\mcitedefaultendpunct}{\mcitedefaultseppunct}\relax
\EndOfBibitem
\end{mcitethebibliography}
